
\def\GR{{\rm I\kern-.23em R}}
\def\GH{{\rm I\kern-.23em H}}
\def\GC{{\rm\kern.24em
  \vrule width.02em height 1.4ex depth-.05ex \kern-.30em C}}
\def\GF{{\rm I\kern-.23em F}}
\def\GN{{\rm I\kern-.21em N}}
\def\GQ{{\rm\kern.24em \vrule width.02em height1.4ex
depth-.05ex
\kern-.30emQ}}
\def\GO{{\rm\kern.24em \vrule width.02em height1.4ex depth-.05ex
\kern-.30em O}}
\def\sgh{{\rm I\kern-.22em H}}
\def\sgc{{\rm I\kern-.32em C}}
\catcode`@=11
\def\eqalignno#1{\displ@y \tabskip\centering
  \halign to\displaywidth{\hfil$\displaystyle{##}$\tabskip\z@skip
    &$\displaystyle{{}##}$\hfil\tabskip\centering
    &\llap{\rm ##}\tabskip\z@skip\crcr
    #1\crcr}}
\catcode`@=12
\def\Rf{\item}

\tolerance=500000
\hoffset=.125in
\voffset=.0625in

{\nopagenumbers
\rightline{IASSNS-HEP-93/32}
\rightline{June 1993~~~~~~}
\bigskip\bigskip
\centerline{\bf Generalized quantum dynamics}
\bigskip\bigskip
\centerline{\it Stephen L. Adler}
\medskip
\centerline{\bf Institute for Advanced Study}
\centerline{\bf Princeton, NJ 08540}
\line{}

\centerline{\bf Abstract}
\baselineskip=20pt
\parindent=.5in
We propose a generalization of Heisenberg picture quantum mechanics in
which a Lagrangian and Hamiltonian dynamics is formulated directly for
dynamical systems on a manifold with non--commuting coordinates, which act
as operators on an underlying Hilbert space. This is accomplished by
defining the Lagrangian and Hamiltonian as the real part of a graded total
trace over the underlying Hilbert space, permitting a consistent definition
of the first variational derivative with respect to a general operator--valued
coordinate.  The Hamiltonian form of the equations is expressed in terms of
a generalized bracket operation, which is conjectured to obey a Jacobi
identity.  The formalism permits the natural implementation of gauge invariance
under operator--valued gauge transformations.  When an operator Hamiltonian
exists as well as a total trace Hamiltonian, as is generally the
case in complex quantum mechanics,
one can make an operator gauge transformation from the Heisenberg to the
Schr\"odinger picture.  When applied to complex quantum
mechanical systems with one bosonic or fermionic degree of freedom, the
formalism gives the usual operator equations of motion, with the canonical
commutation relations emerging as constraints associated with the operator
gauge invariance.  More generally, our methods permit the formulation
of quaternionic quantum field theories with operator--valued gauge invariance,
in which we conjecture that the operator constraints act as a generalization
of the usual canonical commutators.

\vfill\eject}

\pageno=2
\baselineskip=25pt
\centerline{\bf 1. Introduction}

The search for unification of the laws of physics has been closely
intertwined with the discovery of wider classes of symmetries of the
fundamental equations.  Thus, the unification of electricity and magnetism
in the Maxwell equations directly relates to both Abelian gauge invariance
and relativistic kinematics; the advent of the standard model has as its
underpinning the widening of the gauge principle from Abelian to non--Abelian
groups. Current attempts at further unification are based largely on the
use of supersymmetries relating bosonic to fermionic degrees of freedom.

In this paper we explore another possible direction for broadening the
underlying symmetries, by generalizing from c--number to unitary operator
gauge invariance of the fundamental equations.  This approach grew out
of an investigation [1] of quaternionic quantum mechanics and quaternionic
quantum field theory, but the basic concepts can be understood without
going through most of the details of quaternionic Hilbert space, and are
presented here in a self--contained fashion.  We begin, in Sec. 2,  by
developing an operator dynamics based on the idea of a total trace
Lagrangian and Hamiltonian, and show how this can be used to formulate
operator field equations which are covariant under operator--valued gauge
transformations.  We then apply the formalism in Sec. 3 to complex quantum
mechanics, and show that it contains, and generalizes, the usual canonical
formalism.  In Sec. 4, we explain a few simple facts about quaternionic
Hilbert space, and then apply the total trace formalism to construct operator
gauge invariant quaternionic field theories, the properties of which are
discussed.  Some brief concluding remarks are given in Sec. 5.  Further
details of the material presented in this paper can be found  in
the final two chapters of [1].
\vfill\eject


\bigskip
\centerline{\bf 2. Total trace Lagrangian formulation of quantum}
\centerline{\bf dynamics and operator gauge invariance}

The primary tool for achieving invariance under operator--valued gauge
transformations is the concept of a
total trace Lagrangian, which will be developed in general form
in this section. We begin by introducing an underlying complex or quaternionic
Hilbert space, $V_{\GH}$, which we assume to be the direct sum
$$
V_{\GH}=V_{\GH}^+ \oplus V_{\GH}^-
\eqno\rm (1a)
$$
of a Hilbert space $V_\GH^+$ of bosonic states and a Hilbert space
$V_\GH^-$ of fermionic states.  Following Witten [2] we define an
operator $(-1)^F$ which counts fermion number modulo two, that is,
$(-1)^F$ has eigenvalue $+1$ on all states in $V_\GH^+$, and has
eigenvalue $-1$ on all states in $V_\GH^-$.  Using this operator, we
then define a trace operation ${\bf Tr}\,{\cal O}$ for a general operator
${\cal O}$ as follows,
$${\bf Tr}~{\cal O}= {\rm Re}\,Tr\,(-1)^F{\cal O}=
{\rm Re} \sum_n~\langle n|(-1)^F{\cal O}|n\rangle~,
\eqno\rm (1b)
$$
with ${\rm Re}$ the real part, with
$\{|n\rangle\}$ any complete set of states, and with $Tr$ the usual
operator trace or diagonal sum.
The operation
${\bf Tr}$ has the following useful properties:
\item{(i)}
If ${\cal O}={\cal O}^-$ is fermionic, then ${\bf Tr}~{\cal O}^-=0$, since
$$
{\bf Tr}~{\cal O}^- = {\rm Re}~ Tr[(-1)^F{\cal O}^-]={\rm Re}~Tr[{\cal
O}^-(-1)^F]=-{\rm Re}~Tr[(-1)^F{\cal O}^-]=-{\bf Tr}~{\cal O}^-~.
\eqno\rm (1c)
$$
\item{(ii)}
If ${\cal O}={\cal O}^+$ is bosonic, and ${\cal O}^+={\cal O}_{(1)}{\cal
O}_{(2)}$, then ${\cal O}_{(1)}$ and ${\cal O}_{(2)}$ are either both
bosonic or both fermionic, and we have
$$
\eqalign{
{\bf Tr}~{\cal O}_{(1)}{\cal O}_{(2)}
&={\rm Re}~Tr[(-1)^F{\cal O}_{(1)}{\cal
O}_{(2)}]={\rm Re}~Tr[{\cal O}_{(2)}(-1)^F{\cal O}_{(1)}]\cr
&=\pm {\rm Re}~Tr[(-1)^F{\cal O}_{(2)}{\cal O}_{(1)}]
=\pm {\bf Tr}~{\cal O}_{(2)}{\cal O}_{(1)}~,\cr}
\eqno\rm (1d)
$$
with the $+$ sign holding when ${\cal O}_{(1)}$ and ${\cal O}_{(2)}$ are
both bosonic, and the $-$ sign holding when ${\cal O}_{(1)}$ and ${\cal
O}_{(2)}$ are both fermionic.
\item{(iii)}
If ${\cal O}=-{\cal O}^\dagger$ is anti--self--adjoint, then
$$
{\bf Tr}~{\cal O}={\rm Re}~Tr[(-1)^F{\cal O}]={\rm Re}~Tr[(-1)^F
{\cal O}]^\dagger={\rm Re}~Tr[(-1)^F{\cal
O}^\dagger]=-{\bf Tr}~{\cal O}~,
\eqno\rm (1e)
$$
and ${\bf Tr}~{\cal O}$ vanishes.  Correspondingly, if ${\cal O}$ is
self--adjoint, then ${\bf Tr}~{\cal O}$ agrees with
$Tr\,(-1)^F{\cal O}$, which is already
real.
\item{(iv)}
If ${\bf Tr}\displaystyle{\sum_r}~{\cal O}_r\delta q_r=0$ for arbitrary
independent operator variations $\delta q_r$, then each ${\cal O}_r$ must
vanish, while if ${\bf Tr}\displaystyle{\sum_r}~{\cal O}_r \delta q_r=0$
for operator variations $\delta q_r$ restricted to be of either bosonic
or fermionic type, then the part of ${\cal O}_r$ of the same type must
vanish.  The first statement follows from
$$
{\bf Tr}~\sum_r~{\cal O}_r \delta q_r={\rm Re}~\left(\sum_{n,m,r}~\langle
n|(-1)^F{\cal O}_r|m\rangle\langle m|\delta q_r|n\rangle\right)~;
\eqno\rm (1f)
$$
choosing $\langle m|\delta q_r|n\rangle = \overline{\langle n|(-1)^F{\cal
O}_r|m\rangle}$ (with the bar denoting complex or quaternion
conjugation, the latter as defined in Sec.~4 below) gives
$$
{\bf Tr}\sum_r~{\cal O}_r \delta q_r =\sum_{n,m,r}~|\langle n|(-1)^F
{\cal O}_r|m\rangle|^2~,
\eqno\rm (1g)
$$
which can vanish only if $(-1)^F{\cal O}_r=0$, which implies ${\cal
O}_r=0$.  The second statement follows by noting that when $\delta q_r$
is of bosonic or fermionic type, then property (i) implies that
$$
{\bf Tr}~\sum_r~{\cal O}_r\delta q_r={\bf Tr}\sum_r~{\cal
O}_r^{(s)}\delta q_r= {\bf Tr}\sum_r~{\cal O}_r^{(s)}[\delta q_r+\delta
q_r^{(o)}]~,
\eqno\rm (1h)
$$
with ${\cal O}_r^{(s)}$ the part of ${\cal O}_r$ of the same type as
$\delta q_r$ and $\delta q_r^{(o)}$ an arbitrary variation of the
opposite type as $\delta q_r$.  But $\delta q_r+\delta q_r^{(o)}$ is an
unrestricted variation, so the first statement of property (iv) then
implies ${\cal O}_r^{(s)}=0$.

Let now $\{q_r(t)\}$ be a set of time--dependent quantum
variables, which act as operators on the underlying Hilbert space,
with each individual $q_r$ of either bosonic or fermionic
type, and let $\{\dot q_r(t)\}$ be their time derivatives.  We introduce
an operator Lagrangian $L$ which is a polynomial function (or more
generally, a Laurent series expandable function) of the variables
$\{q_r\}$ and $\{\dot q_r\}$,
$$
L=L[\{q_r\},\{\dot q_r\}]~,
\eqno\rm (2a)
$$
and we define the total trace Lagrangian ${\bf L}$ by
$$
{\bf L}[\{q_r\},\{\dot q_r\}]={\bf Tr}~L[\{q_r\},\{\dot q_r\}]~,
\eqno\rm (2b)
$$
and the total trace action ${\bf S}$ by
$$
{\bf S}=\int_{-\infty}^\infty~dt{\bf L}~.
\eqno\rm (2c)
$$
Because of property (i) of ${\bf Tr}$, any fermionic part of $L$ is
automatically projected to zero, so there is no loss in generality in
assuming that $L$ is bosonic.  Similarly, by property (iii) of ${\bf
Tr}$, any anti--self--adjoint part of $L$ is automatically projected to
zero, so we lose no generality by further specifying that $L$ is
self--adjoint.

Let us now examine the consequences of requiring the total trace action
to be stationary under arbitrary {\it operator variations} of the
$\{q_r\}$, subject to the restriction that $\delta q_r$ be of the same
bosonic or fermionic type as $q_r$.  When we vary a given variable
$q_r$, the variation of $L$ consists of a sum of terms of the form
$$
{\cal O}_L \delta q_r{\cal O}_R~,
\eqno\rm (3a)
$$
with ${\cal O}_{L,R}$ operators appearing respectively on the left $(L)$
and right $(R)$ of $\delta q_r$, which in general do not commute with
$\delta q_r$.  Inside the operation ${\bf Tr}$, we can cyclically permute
the factors in Eq.~(3a) to get
$$
{\bf Tr}~{\cal O}_L \delta q_r{\cal O}_R=\pm {\bf Tr}~{\cal O}_R{\cal
O}_L \delta q_r~,
\eqno\rm (3b)
$$
with the $+(-)$ sign corresponding, as in property (ii), to whether ${\cal
O}_R$ is of bosonic (fermionic) type.  Reordering all terms with the
general form of Eq.~(3a) this way, we are able to identify a
well--defined operator $\delta {\bf L}/\delta q_r$, of the same bosonic
or fermionic type as $q_r$, which obeys
$$
\delta {\bf L}={\bf Tr}~{\delta{\bf L}\over\delta q_r}~\delta q_r~.
\eqno\rm (3c)
$$
Similarly, varying one of the $\dot q_r$, we identify a well--defined
operator $\delta{\bf L}/\delta\dot q_r$, again of the same type as
$q_r$, which obeys
$$
\delta {\bf L}={\bf Tr}~{\delta{\bf L}\over \delta \dot q_r}~\delta \dot
q_r~.
\eqno\rm (3d)
$$
When $q_r$ is of bosonic type, the order of the factors within ${\bf
Tr}$ in Eqs.~(3c,d) is irrelevant, but when $q_r$ is of fermionic
type the factor ordering is significant, since by property (ii) of
Eq.~(1d), a minus sign appears when the order of two factors of
fermionic type is reversed.  In many applications, some of the $q_r$ are
either self--adjoint or anti--self--adjoint in character.  If $\delta
q_r$ is further restricted to have the same adjointness
character as
$q_r$, then only the parts of $\delta{\bf L}/\delta q_r$ and $\delta{\bf
L}/\delta\dot q_r$ which have the same (opposite) adjointness character as
a bosonic (fermionic) $q_r$
are well--defined.  It will be assumed henceforth that for those $q_r$
with definite adjointness character, the variational derivatives $\delta{\bf
L}/\delta q_r$ and $\delta{\bf L}/\delta \dot q_r$ denote the operators
of the same (opposite) adjointness character as
a bosonic (fermionic) $q_r$ which obey Eqs.~(3c,d).
We note, finally, that the procedure just described cannot be extended
to higher order variational derivatives.  Since $\delta {\bf L}/\delta
q_r$ is already an operator, a further variation will involve a sum of
terms of the form of Eq.~(3a), in which the $\delta q_r$ (or $\delta
\dot q_r$) factors are sandwiched between operators on left and right
with which they do not commute.  Without the trace, there is now no way
to combine the terms in the sum into a single expression with
infinitesimals on the right, and hence there is no definition of second
variational derivatives analogous to Eqs.~(3c,d).

Let us now impose an action principle, by requiring
$$
\delta {\bf S}=0
\eqno\rm (4a)
$$
under arbitrary same--type operator variations.  Varying all the
arguments $q_r$ and $\dot q_r$ of ${\bf L}$, we get
$$
\delta {\bf S}=\int_{-\infty}^\infty~dt\,\delta{\bf
L}=\int_{-\infty}^\infty~dt\,{\bf Tr}\sum_r~\left({\delta{\bf
L}\over\delta q_r}~\delta q_r+{\delta{\bf L}\over\delta \dot
q_r}~\delta\dot q_r\right)~,
\eqno\rm (4b)
$$
which by an integration by parts becomes
$$
\delta{\bf S}={\bf Tr}~\sum_r~{\delta{\bf L}\over\delta\dot q_r}~\delta
q_r ~\Bigg|_{-\infty}^\infty~+\int_{-\infty}^\infty~dt\,{\bf
Tr}~\sum_r~\left[{\delta{\bf L}\over\delta q_r}-{d\over
dt}~\left({\delta {\bf L}\over \delta \dot q_r}\right)\right]\,\delta q_r~.
\eqno\rm (4c)
$$
So if we take the variations $\delta q_r$ to vanish at $\pm\infty$,
requiring $\delta{\bf S}=0$ gives, by property (iv), the {\it operator}
equations of motion (the generalized Euler--Lagrange equations)
$$
{\delta{\bf L}\over\delta q_r}-{d\over dt}~\left({\delta{\bf L}\over
\delta\dot q_r}\right)=0~.
\eqno\rm (4d)
$$
Corresponding to the Lagrangian form of the equations in Eq.~(4d),
we can set up a Hamiltonian form by the usual method of making a
Legendre transformation.  Defining the momentum operator $p_r$ conjugate
to $q_r$ by
$$
p_r\equiv {\delta{\bf L}\over\delta\dot q_r}~,
\eqno\rm (5a)
$$
we define the total trace Hamiltonian ${\bf H}$ by
$$
{\bf H}={\bf Tr}~\sum_r~p_r\dot q_r-{\bf L}~.
\eqno\rm (5b)
$$
We then have, under general same--type operator variations,
$$
\delta{\bf H}={\bf Tr}~\sum_r~(\delta p_r\dot q_r+p_r\delta\dot
q_r)-{\bf Tr}~\sum_r~\left({\delta{\bf L}\over\delta\dot q_r}~\delta\dot
q_r+{\delta{\bf L}\over\delta q_r}~\delta q_r\right)~,
\eqno\rm (5c)
$$
which substituting Eqs.~(5a) and (4d), and using property (ii),
becomes
$$
\delta{\bf H}={\bf Tr}~\sum_r~(\pm \dot q_r\delta p_r-\dot p_r\delta
q_r)~,
\eqno\rm (5d)
$$
with the $+(-)$ sign chosen according as whether $q_r$ is of bosonic
(fermionic) type.  Equation~(5d) shows that ${\bf H}$ is a total
trace functional of the operators $\{q_r\}$ and $\{p_r\}$,
$$
{\bf H}={\bf H}[\{q_r\},\{p_r\}]~,
\eqno\rm (5e)
$$
with the operator variational derivatives
$$
{\delta{\bf H}\over \delta q_r}=-\dot p_r~,~~~~~~{\delta{\bf
H}\over\delta p_r}=\pm\dot q_r~.
\eqno\rm (5f)
$$
As in the case of the Lagrangian variations, when $q_r$ has a definite
adjointness character, the variations $\delta{\bf H}/\delta q_r$ and
$\delta{\bf H}/\delta p_r$ denote the operators obeying Eq.~(5d)
which have the same (opposite) adjointness character as
a bosonic (fermionic) $q_r$.  We note, finally,
that with $p_r$ defined as in Eq.~(5a), if the Euler--Lagrange
equations are satisfied but arbitrary variations $\delta q_r$ are
permitted at $t=\pm\infty$, then Eq.~(4c) implies that
$$
\delta {\bf S}={\bf Tr}\sum_r~p_r\delta q_r~\bigg|_{-\infty}^\infty~.
\eqno\rm (5g)
$$
This formula and the others involving ${\bf S}$ have obvious
generalizations when the time interval $(-\infty ,\infty)$ is replaced
by $(T_1,T_2)$, for arbitrary finite $T_{1,2}$.

Continuing in analogy with the standard Hamiltonian formalism, let ${\bf
A}[\{q_r\},\{p_r\}]$ and ${\bf B}[\{q_r\},\{p_r\}]$ be any two total
trace functionals of the operator arguments $\{q_r\}$ and $\{p_r\}$, and
let us define their generalized Poisson bracket
$$
{\bf \{A,B\}}\equiv{\bf Tr}\sum_r~(\pm)\left({\delta{\bf A}\over\delta
q_r}~{\delta{\bf B}\over\delta p_r}-{\delta{\bf B}\over\delta
q_r}~{\delta{\bf A}\over\delta p_r}\right)~,
\eqno\rm (6a)
$$
with the $+(-)$ sign again corresponding to $q_r$ bosonic (fermionic).
Then for a general total trace functional ${\bf A}[\{q_r\},\{p_r\}]$ we
have
$$
{\bf \{A,H\}}={\bf Tr}\sum_r~(\pm)\left({\delta{\bf A}\over\delta
q_r}~{\delta{\bf H}\over\delta p_r}-{\delta{\bf H}\over\delta
q_r}~{\delta{\bf A}\over \delta p_r}\right)={\bf Tr}\sum_r~
\left({\delta{\bf A}\over\delta q_r}~\dot q_r+{\delta{\bf A}\over\delta
p_r}~\dot p_r\right)={d\over dt}~{\bf A}~,
\eqno\rm (6b)
$$
and since by construction the generalized Poisson bracket is anti--symmetric
in its arguments,
$$
{\bf\{A,B\}}=-{\bf \{B,A\}}~,
\eqno\rm (6c)
$$
it follows that the time derivative of ${\bf H}$ vanishes,
$$
{d\over dt}~{\bf H}={\bf \{H,H\}}=0~.
\eqno\rm (6d)
$$
An important, and perhaps difficult, open question concerning the
generalized Poisson bracket of Eq.~(6a) is whether it satisfies a
Jacobi identity.  That is, let ${\bf A}[\{q_r\},\{p_r\}],~{\bf
B}[\{q_r\},\{p_r\}]$ and ${\bf C}[\{q_r\},\{p_r\}]$ be any three total
trace functionals of the operator arguments $\{q_r\}, \{p_r\}$, and let
us define the bracket
$$
{\bf [A,B,C]}={\bf\{A,\{B,C\}\}}+{\bf \{C,\{A,B\}\}}+{\bf
\{B,\{C,A\}\}}~,
\eqno\rm (6e)
$$
which is totally antisymmetric in ${\bf A}, {\bf B}$ and ${\bf C}$.  On
the basis of a number of examples, we conjecture that
$$
{\bf [A,B,C]}\equiv 0~,
\eqno\rm (6f)
$$
but we do not have a proof.  It clearly is an important issue to find
either a proof or a counter--example.  If the Jacobi identity is
satisfied, then using Eq.~(6b), the bracket $\{{\bf A},{\bf B}\}$ of any
two constants of the motion ${\bf A}$ and ${\bf B}$ is itself a constant
of the motion.

We now have all the ingredients needed to give a generalized version of
Heisenberg picture quantum mechanics.  States are described by fixed
vectors $|b\rangle \in V_{\GH}^+$ and $|f\rangle\in V_{\GH}^-$, and so
the inner product geometry specified by the set of all inner products
$\{\langle b|b^\prime\rangle\}$ and $\{\langle f|f^\prime\rangle\}$ is
automatically time independent.  The time dependence of the operators
$\{q_r\}$ and $\{p_r\}$ is completely specified by Eq.~(5f), giving
these operators at all times once their form is specified at some
initial time (say $t=0$).  The most general observable ${\cal O}$ will
be a self--adjoint polynomial function (or Laurent expandable function)
of $\{q_r\},\{p_r\}$ and the time $t$,
$$
{\cal O}={\cal O}[\{q_r\},\{p_r\},t]~,
\eqno\rm (7a)
$$
and its time dependence is determined by using the Leibnitz product rule
and Eq.~(5f).  The expectation of ${\cal O}$ in any state
$|b\rangle$ or $|f\rangle$ can be rewritten as a total trace
functional according to
$$
\eqalignno{{\bf\langle{\cal O}\rangle}
&=\left\{{\langle b|{\cal O}|b\rangle ={\bf
Tr}\,P_b{\cal O}~~~~~~\atop \langle f|{\cal O}|f\rangle = -{\bf Tr}\,P_f{\cal
O}~,}\right.\cr
&~~~~~~~P_b=|b\rangle\langle b|~,~~~~~~P_f=|f\rangle\langle f|~.
&(7b)\cr}
$$
This permits us to apply Eq.~(6b), as generalized to the case in
which ${\bf A}$ has an explicit time dependence, giving
$$
{d\over dt}~{\bf \langle{\bf{\cal O}}\rangle}=\langle\partial
{\cal O}/\partial t\rangle+{\bf\{\langle{\cal O}\rangle ,H\}}~.
\eqno\rm (7c)
$$
Transition probabilities can also be reexpressed as total trace
functionals,
$$
|\langle b|b^\prime\rangle|^2={\bf Tr}\,P_bP_{b\prime}={\bf
Tr}\,P_{b^\prime}P_b~,~~~~~~|\langle f|f^\prime\rangle|^2=-{\bf Tr}
P_fP_{f^\prime}=-{\bf Tr}P_{f^\prime}P_f~,
\eqno\rm (7d)
$$
and are time independent by virtue of the time independence of the
projectors $P_b,P_f,\ldots$~.

After this rather lengthy excursion into total trace Lagrangians and
Hamiltonians, we are ready to introduce the concept of operator gauge
invariance.  In its simplest form, an operator gauge transformation
consists of a transformation on the operators $q_r$ of the form
$$
q_r\to U_rq_rU_r^\dagger+\Delta q_r[U_r]~,
\eqno\rm (8a)
$$
with each $U_r$ a unitary operator of bosonic type,
$$
U_rU_r^\dagger=U_r^\dagger U_r=1~,~~~~~~[(-1)^F,U_r]=0~,
\eqno\rm (8b)
$$
and with $\Delta q_r[U_r]$ an inhomogeneous term calculable in terms of
the operator $U_r$.  The Lagrangian $L$ of Eq.~(2a) cannot in
general be constructed to be invariant under the transformation of
Eq.~(8a), but we will find that we can readily construct Lagrangians
$L$ in the form
$$
L=\sum_r~L_r~,
\eqno\rm (8c)
$$
which transform under Eq.~(8a) as
$$
L\to\sum_r~U_rL_rU_r^\dagger~.
\eqno\rm (8d)
$$
As a consequence, although the operator Lagrangian $L$ is not invariant,
the total trace Lagrangian
$$
{\bf L}={\bf Tr}\,L
\eqno\rm (8e)
$$
is invariant under Eq.~(8a),
$$
{\bf L}\to {\bf L}~,
\eqno\rm (8f)
$$
by virtue of the properties of $U_r$ in Eq.~(8b) together with the
cyclic invariance of the trace.  We will also employ a second form of
operator gauge transformation, in which the variables $q_r$ divide into
three groups, the operators $q_r$ in the first group transforming as in
Eq.~(8a), those in the second group transforming as
$$
q_r\to U_r^\prime q_rU_r^{\prime\dagger}+\Delta q_r[U_r^\prime]~,
\eqno\rm (9a)
$$
and those in the third group transforming as
$$
q_r\to U_rq_rU_r^{\prime\dagger}~,
\eqno\rm (9b)
$$
with $U_r,U_r^\prime$ two independent unitary operators of bosonic type.
 We will now find Lagrangians $L$ in the form
$$
L=\sum_r~(L_r+L_r^\prime )~,
\eqno\rm (9c)
$$
which transform under Eqs.~(8a) and (9a,b) as
$$
L\to \sum_r~(U_rL_rU_r^\dagger+U_r^\prime L_r^\prime
U_r^{\prime\dagger})~.
\eqno\rm (9d)
$$
Again, while the operator Lagrangian $L$ of Eq.~(9c) is not operator
gauge invariant, the corresponding total trace Lagrangian ${\bf L}$ is
operator gauge invariant.

We now make a number of remarks concerning the structure and properties
of operator gauge invariant total trace Lagrangians.
\item{(1)}
We have written the equations of this section with $r$ a discrete index,
but in many of the applications described in the next two sections, $r$
will be replaced by a continuum coordinate $\vec x$.
\item{(2)}
Varying Eqs.~(8a), (9a) and (9b) with respect to $q_r$, the
inhomogeneous term drops out, and we get respectively
$$
\eqalignno{\delta q_r
&\to U_r \delta q_rU_r^\dagger~~~~~~{(\rm first~ group)}~,\cr
\delta q_r &\to U_r^\prime \delta q_r U_r^{\prime\dagger}~~~~~~{(\rm
second~ group)}~,\cr
\delta q_r &\to U_r \delta
q_rU_r^{\prime\dagger}~~~~~~{\rm(third~group)}~.&(9e)\cr}
$$
Hence when ${\bf L}$ is operator gauge invariant, the Eulerian
derivative appearing in Eq.~(4c) transforms as
$$
\eqalignno{E_r
&\to U_rE_rU_r^\dagger~~~~~~{\rm(first~group)}~,\cr
E_r &\to U_r^\prime E_rU_r^{\prime\dagger}~~~~~~{\rm(second~group)}~,\cr
E_r &\to U_r^\prime E_rU_r^\dagger~~~~~~{\rm(third~group)}~,\cr
E_r &\equiv {\delta {\bf L}\over \delta q_r}-{d\over dt}~{\delta{\bf
L}\over\delta\dot q_r}~, &(9f)\cr}
$$
and the Euler--Lagrange equations $E_r=0$ are operator gauge covariant.
\item{(3)}
A total trace version of the familiar Noether theorem of
classical mechanics can be derived, as follows.  Let $\delta\Lambda(t)$
be an infinitesimal operator parameterizing a set of operator variations
$\delta q_r$ of the variables $q_r$, and let us assume that $\delta{\bf
L}$ only involves $\delta\Lambda(t)$ and $\delta\dot\Lambda(t)$, but not
$\delta\ddot\Lambda(t)$ or higher time derivatives (the Lagrangians
studied in the next two sections all have this feature).  Then we have
$$
\eqalignno{\delta{\bf L}
&={\bf Tr}\left({\delta{\bf
L}\over\delta\Lambda}~\delta\Lambda+{\delta{\bf
L}\over\delta\dot\Lambda}~\delta\dot\Lambda\right)~,\cr
\delta{\bf S}&={\bf Tr}~{\delta{\bf
L}\over\delta\dot\Lambda}~\delta\Lambda~\Bigg|_{-\infty}^\infty+
\int_{-\infty}^\infty~dt\,{\bf Tr}\left(\left[{\delta{\bf
L}\over\delta\Lambda}-{d\over dt}~\left({\delta{\bf
L}\over\delta\dot\Lambda}\right)\right]\delta\Lambda\right)~.
&(10a)\cr}
$$
Assuming now that $\delta\Lambda$ vanishes rapidly enough at
$t=\pm\infty$ so that all the $\delta q_r$ vanish there, and that the
generalized Euler--Lagrange equations are satisfied for all times, then
both $\delta{\bf S}$ and the surface terms in Eq.~(10a) vanish, and
independence of $\delta\Lambda (t)$ at different times implies
$$
0={\bf Tr}\,\left(\left[{\delta{\bf L}\over\delta\Lambda} -{d\over
dt}~\left({\delta{\bf
L}\over\delta\dot\Lambda}\right)\right]\delta\Lambda\right)~.
\eqno\rm (10b)
$$
Suppose now that ${\bf L}$ is left invariant under the variations
$\delta q_r$ parameterized by $\delta\Lambda$.  Then $\delta{\bf
L}/\delta\Lambda =0$, and Eq.~(10b) simplifies to
$$
{\bf Tr}\,\left({dQ_\Lambda\over dt}~\delta\Lambda\right)=0~,
\eqno\rm (10c)
$$
with $Q_\Lambda$ the ``charge'' defined by
$$
Q_\Lambda={\delta{\bf L}\over\delta\dot\Lambda}~.
\eqno\rm (10d)
$$
If ${\bf L}$ is invariant for arbitrary time--independent anti--self--adjoint
operators $\delta\Lambda$, then Eq.~(10c) implies
the operator statement
$$
{dQ_\Lambda\over dt}=0~,
\eqno\rm (10e)
$$
with $Q_\Lambda$ anti--self--adjoint.  (Operator gauge transformations
obey this condition trivially, with $Q_\Lambda \equiv 0$, since ${\bf L}$
is invariant for arbitrary time--dependent anti--self--adjoint
$\delta\Lambda$.)
On the other hand, in the case of Poincar\'e transformations,
$\delta\Lambda$ is a $c$--number describing an infinitesimal translation
or proper Lorentz transformation of the coordinates, and Eq.~(10c)
then only implies the total trace
relation
$$
{d\over dt}~{\bf Tr}~Q_\Lambda =0~,
\eqno\rm (10f)
$$
of which Eq.~(6d) is a particular example.
\item{}
{}~~~~~An important special case is that in which $\delta\Lambda (t)$
parameterizes a linear transformation of the coordinates of the form
$$
\delta q_r=\delta\Lambda(t)\sum_s~G_{rs}q_s~,
\eqno\rm (11a)
$$
with $G_{rs}$ independent of time and of the $q$'s and $\dot q$'s.  Then
we have
$$
\delta\dot
q_r=\delta\dot\Lambda(t)\sum_s~G_{rs}q_s+\delta\Lambda(t)\sum_s~G_{rs}\dot
q_{s}~,
\eqno\rm (11b)
$$
which together with Eq.~(3d) implies that
$$
Q_G={\delta{\bf
L}\over\delta\dot\Lambda}=\pm\sum_{rs}~G_{rs}q_s{\delta{\bf
L}\over\delta\dot q_r}~,
\eqno\rm (11c)
$$
again with the $+(-)$ sign chosen according as whether $q_r$ is of
bosonic (fermionic) type.  Substituting Eq.~(5a) into Eq.~(11c)
and taking the trace, we thus get
$$
{\bf Tr}\,Q_G={\bf Tr}\left(\sum_{rs}~p_rG_{rs}q_s\right)~.
\eqno\rm (11d)
$$
Let now $Q_H$ be the charge associated with a second linear
transformation in which $G_{rs}$ is replaced by $H_{rs}$; then for the
generalized Poisson bracket of ${\bf Tr}\,Q_G$ with ${\bf Tr}\,Q_H$, we
find
$$
\eqalignno{\{{\bf Tr}\,Q_G,{\bf Tr}\,Q_H\}
&={\bf Tr}
\left(~\sum_{rst}~(\pm)(p_rG_{rs}(\pm)H_{st}q_t-p_rH_{rs}
(\pm)G_{st}q_t)\right)\cr
&={\bf Tr}\,\left(\sum_{rt}~p_r\sum_s~(G_{rs}H_{st}-H_{rs}G_{st})q_t\right)
={\bf Tr}\,Q_{[G,H]}~.~~~ &(11e)\cr}
$$
Hence if a set of matrices $G,H,\ldots$ used to generate linear
transformations of the $q$'s obeys a Lie algebra, then the corresponding
functionals ${\bf Tr}\,Q_G,~{\bf Tr}\,Q_H,\ldots$ obey the same Lie
algebra under the generalized Poisson bracket.  This shows, for example,
that in a Poincar\'e invariant theory defined by a total trace action,
the total trace functionals defining the Poincar\'e generators will obey
the Poincar\'e algebra under the bracket operation of Eq.~(6a).
\item{(4)}
When an operator gauge invariance is present, the problem of identifying
physical observables becomes more subtle.  Since observables should
correspond to invariant geometric features of the quantum dynamics, only
operator gauge invariant quantities can be observables.  Thus, the
expectation values of Eq.~(7b), which are not operator gauge
invariant for general states $|b\rangle ,|f\rangle$, are in general not
observables.  One way to form observable quantities is to construct
total trace functionals, similar to ${\bf L}$ but involving higher degree
polynomials in the operators $q_r$, which are operator gauge invariant.
Clearly, an infinite number of such observables can be constructed.  A
second way to form observable quantities is to focus on particular
operators ${\cal O}_{rs}$ transforming under operator gauge
transformations as
$$
{\cal O}_{rs} \to U_r{\cal O}_{rs}U_s^\dagger~,
\eqno\rm (12a)
$$
and on the co--transforming bases of states
$|b_r^{(n)}\rangle,|b_s^{(n)}\rangle,|f_r^{(n)}\rangle,|f_s^{(n)}\rangle,
n=1,2,\ldots$, which transform as
$$
\eqalignno{|b_r^{(n)}\rangle
&\to U_r|b_r^{(n)}\rangle~,~~~~~~|b_s^{(n)}\rangle \to
U_s|b_s^{(n)}\rangle~,\cr
|f_r^{(n)}\rangle &\to U_r|f_r^{(n)}\rangle~,~~~~~~|f_s^{(n)}\rangle \to
U_s|f_s^{(n)}\rangle~. &(12b)\cr}
$$
Then the special class of matrix elements
$$
\langle b_r^{(n)}|{\cal O}_{rs}|b_s^{(m)}\rangle~,~~~~~~\langle f_r^{(n)}|
{\cal O}_{rs}|f_s^{(m)}\rangle~,
\eqno\rm (12c)
$$
are operator gauge invariant, and hence
observables.
\item{(5)}
As we have seen, in theories constructed from a total trace Lagrangian,
the Hamiltonian dynamics is governed by the operator variational
equations of Eq.~(5f), which are generated by the total trace
Hamiltonian ${\bf H}$.  There appears to be no special reason for this
dynamics to be unitary, that is, in general there is no reason to expect
that there should be a unitary time evolution operator $U(t,0)$, such
that for all $q_r, p_r$ and all times $t$, the dynamics of Eq.~(5f)
is equivalent to
$$
q_r(t)=U^\dagger(t,0)q_r(0)U(t,0)~,~~~~~~p_r(t)=U^\dagger(t,0)p_r(0)U(t,0)~.
\eqno\rm (13a)
$$
An equivalent statement in infinitesimal form is
that in general there is no reason to expect that there should be an
anti--self--adjoint operator Hamiltonian $\tilde H(t)$, such that for
all $q_r,p_r$ and all times $t$, the dynamics of Eq.~(5f) is
equivalent to
$$
\dot q_r=[\tilde H(t),q_r]~,~~~~~~\dot p_r=[\tilde H(t),p_r]~.
\eqno\rm (13b)
$$
It follows from these statements that total trace Lagrangian dynamics
is, potentially, an even more general form of quantum mechanics than standard,
operator Hamiltonian--based, quantum mechanics.
\item{}
{}~~~~~Suppose, however, that the dynamics is such that a $U(t,0)$ and an
$\tilde H(t)$ obeying Eqs.~(13a,b) {\it do} exist, for a theory with
operator gauge invariance.  Then we can make an operator gauge
transformation
$$
q_r(t)\to U(t,0)q_r(t)U^\dagger(t,0)\equiv q_{rS}(t)~,~~~~~~p_r(t)\to
U(t,0)p_r(t)U^\dagger(t,0)\equiv p_{rS}(t)~,
\eqno\rm (13c)
$$
with a corresponding transformation for co--transforming states
$|b\rangle,|f\rangle$,
$$
|b\rangle\to U(t,0)|b\rangle\equiv |b_S(t)\rangle~,~~~~~~|f\rangle\to
U(t,0)|f\rangle\equiv |f_S(t)\rangle~,
\eqno\rm (13d)
$$
thus defining ``Schr\"odinger picture'' operators $q_{rS}(t),p_{rS}(t)$
and states $|b_S(t)\rangle$ and $|f_S(t)\rangle$.  By construction, the
operators $q_{rS}(t)$ and $p_{rS}(t)$ are time independent,
$$
q_{rS}(t)=q_{rS}(0)~,~~~~~~p_{rS}(t)=p_{rS}(0)~,
\eqno\rm (13e)
$$
while the states $|b_S(t)\rangle$ and $|f_S(t)\rangle$ obey the
Schr\"odinger time--development equation
$$
{d\over dt}~|b_S(t)\rangle = - \tilde H(t)|b_S(t)\rangle~,~~~~~~{d\over
dt}~|f_S(t)\rangle =-\tilde H(t)|f_S(t)\rangle~.
\eqno\rm (13f)
$$
{}From this perspective, Schr\"odinger picture quantum mechanics appears
as a rather special case of the more general quantum dynamics described
by operator gauge invariant total trace Lagrangians.
\item{}
{}~~~~~It clearly is of great importance to determine under what circumstances
the operator time development equations of Eq.~(5f) are equivalent
to a unitary evolution as in Eqs.~(13a,b).  It seems likely that this
equivalence always holds in complex quantum mechanics, since there the
standard canonical quantization rules give a constructive procedure for
going from the Lagrangian $L$, interpreted now as a classical
Lagrangian, to an operator Hamiltonian obeying Eq.~(13b).  In the
case of quaternionic quantum mechanics the situation is far from clear,
and we leave the problem of determining whether and when Eq.~(5f) is
equivalent to Eqs.~(13a,b) as an important open question.

\item{(6)}
As formulated up to this point, total trace quantum dynamics applies for
arbitrary operator properties of the dynamical variables
$\{q_r\},\{p_r\}$ at some initial time (say $t=0$), from which the
Hamiltonian equations of motion can be integrated forwards to $t>0$ (and
backwards to $t<0$ as well).  Suppose now that for some subset of the
variables $\{q_R\}$, the Lagrangian ${\bf L}$ is independent of the time
derivatives $\{\dot q_R\}$.  Then the corresponding canonical momenta
$\{p_R\}$ vanish identically,
$$
p_R={\delta{\bf L}\over \delta\dot q_R}\equiv 0~,
\eqno\rm (14a)
$$
and the Euler--Lagrange equations for these variables degenerate to the
constraints
$$
{\delta{\bf L}\over \delta q_R}=0~.
\eqno\rm (14b)
$$
We are now dealing with a constrained Hamiltonian system, for which an
operator generalization of the standard Dirac treatment of
constrained systems will be
needed. This may involve some subtleties, since the general Dirac
prescription employs second variational derivatives of the Lagrangian,
which, as discussed above, are not well--defined in the operator case.
The situation in
which some of the canonical momenta vanish identically is of course not
the most general form of a constrained system, but it is precisely what
occurs when a gauge invariance is present.  We assume, in analogy with
the standard Yang--Mills case, that the correct procedure for operator
gauge invariant systems will be to adjoin to the operator constraints of
Eqs.~(14a,b) an equal number of operator gauge--fixing conditions,
which break the operator gauge invariance.   {\it We
conjecture that the constraints of Eqs.~(14a,b), together with the
operator gauge--fixing conditions, provide the minimum specification of
operator properties of the $\{q_r\}$ and $\{p_r\}$ which are needed for
a consistent theory.}  We will see, however, that there are examples in
which it is possible to add further constraints beyond this minimum and
still preserve consistency with the operator equations of motion.  Finding
a suitable operator generalization of the
Dirac procedure, at least in the case of operator gauge invariant
systems, and determining the precise conditions needed for operator
specification, are again important open problems.

\item{(7)}
Finally, some historical remarks.  The concepts of
operator--valued gauge transformations and a total trace action were
introduced, without the $(-1)^F$ factor and real part of Eq.~(1b),
by Adler [3,4,5]
in the context of a theory termed ``algebraic chromodynamics.''   The
$(-1)^F$ factor was used by Witten [2] to define a topological index,
which in our notation is simply ${\bf Tr}\,1$.  The necessity for
including the real part in the definition of a trace
for quaternionic Hilbert space was pointed out by Finkelstein, Jauch,
and Speiser [6].  A general suggestion of
operator--valued gauge transformations was also made by Mackey
[7], with an implementation [8] which formulates the
equivalence class, under unitary operator transformations, of complex
Galilean invariant Hamiltonians. Since what we have done above is to set up a
dynamics on a manifold with non--commuting coordinates $\{q_r\}$, the
discussion of this and the subsequent two sections appears to be
related to the non--commutative geometry program of A. Connes [9,10].
There are close
analogies between the identification of observables in operator gauge
invariant theories, and the identification of observables in general
relativity and in conventional Yang--Mills gauge theories.  Operators
transforming as in Eq.~(12a) are analogs of bitensor quantities in
classical general relativity and of path ordered
integrals in gauge theories.  The suggestion that the generalization
from special to general relativity should have an operator analog in the
generalization from complex to quaternionic quantum mechanics, was first
made by Finkelstein, Jauch, Schiminovich, and Speiser [11].

Because the analysis of this section has dealt with the general case, it
has of necessity been rather abstract.  Concrete illustrations of
operator gauge invariant systems, in complex and quaternionic quantum
mechanics and quantum field theory, are given in the next two sections.
\vfill\eject
\centerline{\bf 3. Operator gauge invariant total trace Lagrangian formulation}
\centerline{\bf of complex quantum mechanics}

We proceed in this section to illustrate the general formalism which we
have just set up, in the familiar context of complex quantum mechanics
and quantum field theory.  In complex quantum mechanics, the left--acting
operator $I$ defined by
$$
I=\sum_n~|n\rangle\,i\,\langle n|
\eqno\rm (15)
$$
commutes with all operators. (This permits the identification of
$I$ with $i1$, with $1$ the unit operator and $i$ the
right--acting imaginary unit, a notation which is standard in the
complex quantum mechanics literature but which we shall not follow here.
A discussion of right--acting versus left--acting algebras is given in
Sec.~4 below.)
Hence the complex specialization of operator gauge invariance consists
in assuming that the operators $U_r,U_r^\prime$ of Eqs.~(8a) and
(9a,b) commute with $I$, which permits the inclusion of explicit
factors of $I$ in the construction of operator gauge invariant total
trace Lagrangians.

As our first example, we consider a single self--adjoint bosonic
coordinate $q(t)$ obeying Galilean invariant dynamics.  The conventional
Lagrangian for this model (with the mass taken as unity for convenience)
is
$$
L_q^c={1\over 2}~\dot q^2+{1\over 2}~\{\dot q,A(q)\}_+-V(q)~,
\eqno\rm (16a)
$$
with $A(q)$ and $V(q)$ self--adjoint functions of $q$ and $\{~,~\}_+$
the conventional anti--commutator; this Lagrangian
clearly does not have any simple transformation properties under the
operator transformation
$$
q\to UqU^\dagger~,~~~~~~UU^\dagger=U^\dagger U=1~.
\eqno\rm (16b)
$$
To achieve covariance, we follow the standard procedure of replacing the
ordinary time derivative $\partial_0=\partial/\partial t$ by a covariant
derivative $\hat D_0$, defined as
$$
\hat D_0 q\equiv {\partial q\over\partial t}+[B_0,q]~,
\eqno\rm (16c)
$$
with $B_0$ an anti--self--adjoint operator gauge potential.  Under the
transformation of Eq.~(16b), $B_0$ is taken to transform as
$$
B_0\to UB_0U^\dagger -{\partial U\over\partial t}~U^\dagger =
UB_0U^\dagger +U~{\partial\over \partial t}~U^\dagger~,
\eqno\rm (16d)
$$
as a consequence of which $\hat D_0q$ transforms as
$$
\eqalignno{
&\hat D_0q\to{\partial\over\partial
t}~(UqU^\dagger)+\left(UB_0U^\dagger-{\partial U\over\partial
t}~U^\dagger\right) UqU^\dagger-UqU^\dagger\left(UB_0U^\dagger+
U~{\partial\over\partial t}~U^\dagger\right)\cr
&~~~~~~~~=U\left({\partial q\over\partial
t}+[B_0,q]\right)U^\dagger=U(\hat D_0q)U^\dagger~. &(16e)\cr}
$$
Hence if we redefine the Lagrangian of Eq.~(16a) as
$$
L_q={1\over 2}~(\hat D_0q)^2+{1\over 2}~\{\hat D_0q,A(q)\}_+-V(q)~,
\eqno\rm (17a)
$$
then under operator gauge transformations $L_q$ transforms
covariantly,
$$
L_q\to UL_qU^\dagger~,
\eqno\rm (17b)
$$
and the corresponding total trace Lagrangian ${\bf L}_q$ and action
${\bf S}_q$ defined by
$$
{\bf L}_q={\bf Tr}\,L_q~,~~~~~~{\bf S}_q=\int_{-\infty}^\infty~dt\,{\bf
L}_q~,
\eqno\rm (17c)
$$
are invariant.

We have now achieved operator gauge invariance at the price of
introducing an extra dynamical variable $B_0$.  We must next investigate
the structure of possible Lagrangians ${\bf L}_{B_0}$ to govern the
dynamics of $B_0$.  The usual Yang--Mills type field strength $F_{00}$
vanishes identically,
$$
F_{00}=\partial_0B_0-\partial_0B_0+[B_0,B_0]=0~,
\eqno\rm (18a)
$$
and so cannot be used to construct a Lagrangian for $B_0$.
However, there is one additional Lagrangian which can be formed from
$B_0$,
$$
{\bf L}_{B_0}={\bf Tr}\,(IB_0)~,
\eqno\rm (18b)
$$
with the inclusion of a factor of $I$ necessitated by the fact that
$B_0$ is anti--self--adjoint.  To see that Eq.~(18b) defines a
satisfactory Lagrangian, we note that it suffices to check its behavior
under infinitesimal operator gauge transformations of the form
$$
U=1+\delta\Lambda~,~~~~~~\delta\Lambda =-\delta\Lambda^\dagger~,
\eqno\rm (18c)
$$
under which the first order variation of $B_0$ is
$$
\delta B_0=[\delta\Lambda,B_0]-{\partial\delta\Lambda\over\partial
t}=-\hat D_0\delta\Lambda~.
\eqno\rm (18d)
$$
{}From Eqs.~(18b,d) we find
$$
\delta{\bf L}_{B_0}={\bf Tr}\,(I\delta B_0)={\bf
Tr}\,\left([B_0,I]\delta\Lambda\right)-{\partial\over\partial t}~{\bf
Tr}\,(I\delta\Lambda)=-{\partial\over\partial t}\,{\bf
Tr}\,(I\delta\Lambda)~,
\eqno\rm (18e)
$$
which is a time derivative, and as a consequence the total trace action
${\bf S}_{B_0}$ defined by
$$
{\bf S}_{B_0}=\int_{-\infty}^\infty~dt\,{\bf L}_{B_0}
\eqno\rm (19a)
$$
is invariant under Eq.~(18c) when $\delta\Lambda$ vanishes at
$t=\pm\infty$ [or more generally, when
$\delta\Lambda(\infty)=\delta\Lambda(-\infty)$],
$$
\delta {\bf S}_{B_0}=0~.
\eqno\rm (19b)
$$
This argument of
course does not imply that ${\bf S}_{B_0}$ is invariant under the global
transformation of Eq.~(16d), which can produce changes in the topological
sector; rather, what we have shown is that ${\bf S}_{B_0}$ defines a
form of topological action, which is constant within each distinct
topological sector defined under operator valued gauge transformation.

An alternative argument for the invariance of ${\bf S}_{B_0}$
proceeds directly from Eq.~(16d), which implies that
$$
{\bf Tr}\,(IB_0)\to{\bf Tr}\,\left[I\left(UB_0U^\dagger-{\partial
U\over\partial t}~U^\dagger\right)\right]={\bf Tr}\,(IB_0)-{\bf
Tr}\,\left(I~{\partial U\over\partial t}~U^\dagger\right)~.
\eqno\rm(19c)
$$
Writing $U=e^{\Lambda(t)}$,  with $\Lambda=-\Lambda^\dagger$, and using
the operator identity (which can be verified by power series expansion
and term--by--term integration)
$$
{\partial\over\partial
t}~e^\Lambda=\int_0^1~ds\,e^{s\Lambda}{\partial\Lambda\over\partial
t}~e^{(1-s)\Lambda}~,
\eqno\rm(19d)
$$
we have
$$
{\bf Tr}\,\left(I{\partial U\over\partial t}~U^\dagger\right)={\bf
Tr}\,\left(I\int_0^1~ds\,e^{s\Lambda}{\partial\Lambda\over\partial
t}~e^{(1-s)\Lambda}e^{-\Lambda}\right)={\bf
Tr}\,\left(I\int_0^1~ds{\partial\Lambda\over\partial t}\right)={\bf
Tr}\,\left(I{\partial\Lambda\over\partial t}\right)~.
\eqno\rm(19e)
$$
Hence ${\bf S}_{B_0}$ is invariant whenever ${\bf
Tr}\,[I\Lambda(\infty)]={\bf Tr}\,[I\Lambda(-\infty)]$.  This argument,
as well as the infinitesimal one given above, extends immediately to
the case in
which $B_0$ is replaced by a space--time component of a four vector
gauge potential and
$\partial/\partial t$ is replaced by a space--time
derivative.

Let us now examine the dynamics following from a general linear
combination of the actions ${\bf S}_q$ and ${\bf S}_{B_0}$.  Forming the
total Lagrangian ${\bf L}$ and action ${\bf S}$,
$$
{\bf L}={\bf L}_q-\lambda_0{\bf L}_{B_0}~,~~~~~~{\bf S}={\bf
S}_q-\lambda_0{\bf S}_{B_0}~,
\eqno\rm (20a)
$$
with $\lambda_0$ a constant, and taking general operator variations, we
get (with repeated use of the cyclic property of the trace)
$$
\eqalign{\delta{\bf L}
&={\bf Tr}\,\left\{[ \hat D_0q +A(q)]\delta(\hat D_0q)+(\hat D_0 q)\delta A(q)
-\delta V(q)-\lambda_0I\delta B_0\right\}\cr
&={\bf Tr}\,\left\{[\hat D_0q+A(q)](\delta \dot q+[\delta
B_0,q]+[B_0,\delta q])+F(q,\hat D_0q)\delta q-\lambda_0I\delta
B_0\right\}\cr
&={\bf Tr}\,\left\{[\hat D_0q+A(q)]\delta\dot q+\left([\hat
D_0q+A(q),B_0]+F(q,\hat D_0q)\right)\delta q\right.\cr
&~~~~~~~~~~~~~\left.+\left([q,\hat D_0q+A(q)]
-\lambda_0I\right)\delta B_0\right\}~, \cr}
\eqno\rm (20b)
$$
where we have defined a generalized force term $F(q,\hat D_0q)$ by
$$
{\bf Tr}\,[(\hat D_0q)\delta A(q)-\delta V(q)]={\bf Tr}\,[F(q,\hat
D_0q)\delta q]~.
\eqno\rm (20c)
$$
Hence
$$
{\delta{\bf L}\over\delta \dot q}=\hat D_0q+A(q)~,~~~~~~{\delta{\bf
L}\over\delta q}=[\hat D_0q+A(q),B_0]+F(q,\hat D_0q)~,
\eqno\rm (20d)
$$
$$
{\delta{\bf L}\over\delta\dot B_0}=0~,~~~~~~{\delta{\bf L}\over\delta
B_0}=[q,\hat D_0q+A(q)]-\lambda_0I~,
\eqno\rm (20e)
$$
and so the Euler--Lagrange equations following from $\delta{\bf S}=0$
consist of a dynamical equation for $q(t)$,
$$
\hat D_0[\hat D_0q+A(q)]={d\over dt}~[\hat D_0q+A(q)]+[B_0,\hat
D_0q+A(q)]=F(q,\hat D_0q)~,
\eqno\rm (20f)
$$
together with a constraint
$$
[q,\hat D_0q+A(q)]=\lambda_0I~,
\eqno\rm (20g)
$$
both of which are covariant under the operator gauge transformations of
Eqs.~(16b--e).  When we rewrite the dynamics in total trace Hamiltonian
form, we identify the canonical momentum conjugate to $q$ as
$$
p={\delta{\bf L}\over\delta\dot q}=\hat D_0q+A(q)~,
\eqno\rm (21a)
$$
and so the constraint of Eq.~(20g) reads
$$
[q,p]=\lambda_0I~,
\eqno\rm (21b)
$$
and is just the standard canonical commutator with the identification
$\lambda_0=\hbar$($=1$ in microscopic units).  Note that because of the
real part in the definition of ${\bf Tr}$, we have ${\bf Tr}\,I=0$, and so
taking ${\bf Tr}$ of the left-- and right--hand sides of Eq.~(21b) gives a
consistent equation for non--zero $\lambda_0$,
$$
0={\bf Tr}\,[q,p]=\lambda_0{\bf Tr}\,I=\lambda_00~.
\eqno\rm (21c)
$$
Carrying out the Legendre
transform of Eq.~(5b) we get
$$
\eqalignno{{\bf H}
&={\bf Tr}\,\left({\delta{\bf L}\over\delta\dot q}\,\dot q\right)-{\bf L}\cr
&={\bf Tr}\,\left\{[\hat D_0q+A(q)](\hat D_0q-[B_0,q])-{1\over 2}~(\hat
D_0q)^2-(\hat D_0q)A(q)+V(q)+\lambda_0IB_0\right\}\cr
&={\bf Tr}\,\left\{{1\over 2}~(\hat D_0q)^2+V(q)+(\lambda_0I-[q,\hat
D_0q+A(q)])B_0\right\}\cr
&={\bf Tr}\,\left\{{1\over
2}~[p-A(q)]^2+V(q)+(\lambda_0I-[q,p])B_0\right\}~, &(21d)\cr}
$$
which in the ``Hamiltonian
gauge''
$$
B_0=0
\eqno\rm (21e)
$$
simplifies to
$$
{\bf H}={\bf Tr}\,\{{1\over 2}~[p-A(q)]^2+V(q)\}~.
\eqno\rm (22a)
$$
Taking the operator variation of Eq.~(22a), and recalling the
definition of Eq.~(20c), we get
$$
{\bf \delta}{\bf H}={\bf Tr}\,\{[p-A(q)][\delta p-\delta A(q)]+\delta
V(q)\}={\bf Tr}\,\{[p-A(q)]\delta p-F(q,p-A(q))\delta q\}~.
\eqno\rm (22b)
$$
Hence the total trace Hamiltonian equations of motion are
$$
\dot q={\delta {\bf H}\over \delta p}=p-A(q)~,~~~~~~\dot p=-{\delta{\bf
H}\over\delta
q}=F(q,p-A(q))~,
\eqno\rm (22c)
$$
which agree with Eqs.~(16c), (20f) and (21a) when these are
specialized to the gauge $B_0=0$.

We see, then, that the Heisenberg picture equations of motion and the
canonical commutation relation for a Galilean particle both emerge from
the operator gauge invariant, total trace Lagrangian formalism.  The
derivations just given are expressed in operator terms throughout; at no
point did we introduce a classical Lagrangian and its ``quantization.''
In complex quantum mechanics, the conventional canonical quantization
route is of course still valid, and implies that there is an operator
Hamiltonian $H$ given by the Weyl ordering of the corresponding
classical Hamiltonian. For the model under study, we have
$$
H=\{{1\over 2}~[p-A(q)]^2+V(q)\}_W~,
\eqno\rm (23a)
$$
where the Weyl ordering subscript $W$ implies symmetrization of $p$ with
respect to the factors of $q$ in each term of $A(q)$, e.g.,
$$
\{pq^n\}_W=\{q^np\}_W={1\over
n+1}~(q^np+q^{n-1}pq+q^{n-2}pq^2+\ldots+q^2pq^{n-2}+qpq^{n-1}+pq^n)~.
\eqno\rm (23b)
$$
With $[q,p]=\lambda_0I$, we then find that
$$
{1\over I\lambda_0}~[q,H]=p-A(q)~,~~~~~~{1\over -I
\lambda_0}~[p,H]=-F(q,p-A(q))~,
\eqno\rm (23c)
$$
since, as may be verified by some algebra, the Weyl ordering of $H$
leads to the same factor ordering in the
force term $F$ as is obtained, via the cyclic property of the trace,
from the operator variational definition of $F$ in Eq.~(20c).  Hence
the equations of motion of Eq.~(22c) are equivalent to
$$
\dot q=\lambda_0^{-1}[IH,q]~,~~~~~~\dot p=\lambda_0^{-1}[IH,p]~,
\eqno\rm (23d)
$$
and so in the terminology of the preceding section, the dynamics is
unitary.  This permits us to transform from the Heisenberg picture to
the Schr\"odinger picture, in which the operators are time independent
and the co--transforming states carry the quantum dynamics.

In the example just given, all operators are bosonic, and so the
$(-1)^F$ factor in the definition of ${\bf Tr}$ does not come into play.
 As our second example, we consider a single non--interacting fermion
degree of freedom with mass $m$, described by the conventional
Lagrangian
$$
L_\psi^c={I\over
2}~(\psi^\dagger\dot\psi-\dot\psi^\dagger\psi)-m\psi^\dagger\psi~.
\eqno\rm (24a)
$$
The Lagrangian of Eq.~(24a) again does not transform simply under the
operator transformation
$$
\psi\to U\psi U^\dagger~,~~~~~~UU^\dagger=U^\dagger U=1~,
\eqno\rm (24b)
$$
but as in the bosonic example, we can achieve covariance by replacing
the time derivative by the covariant derivative $\hat D_0$,
$$
\hat D_0\psi\equiv {\partial\psi\over\partial t}+[B_0,\psi]~,~~~~~~
(\hat D_0\psi)^\dagger=\hat
D_0\psi^\dagger={\partial\psi^\dagger\over\partial t}+[B_0,
\psi^\dagger]~.
\eqno\rm (24c)
$$
With $B_0$ transforming as in Eq.~(16d), $\hat D_0\psi$ transforms as
$$
\hat D_0\psi\to U(\hat D_0\psi)U^\dagger~,~~~~~~\hat D_0\psi^\dagger\to U(\hat
D_0\psi^\dagger)U^\dagger~,
\eqno\rm (24d)
$$
and so the redefined Lagrangian
$$
L_\psi={I\over 2}~[\psi^\dagger \hat D_0\psi-(\hat
D_0\psi^\dagger)\psi]-m\psi^\dagger\psi
\eqno\rm (24e)
$$
transforms covariantly,
$$
L_\psi\to UL_\psi U^\dagger~.
\eqno\rm (24f)
$$
The corresponding total trace Lagrangian ${\bf L}_\psi$ and action ${\bf
S}_\psi$~,
$$
{\bf L}_\psi={\bf Tr}\,L_\psi,~~~~~~{\bf
S}_\psi=\int_{-\infty}^\infty~dt~{\bf L}_\psi~,
\eqno\rm (24g)
$$
are then invariant.

Proceeding as in our first example, we examine the dynamics following
from a general linear combination of the actions ${\bf S}_\psi$ and
${\bf S}_{B_0}$.  Writing
$$
{\bf L}={\bf L}_\psi-\lambda_0{\bf L}_{B_0}~,~~~~~~{\bf S}={\bf
S}_\psi-\lambda_0{\bf S}_{B_0}~,
\eqno\rm (25a)
$$
and taking general operator variations of $\psi$ and $B_0$, with
$\delta\psi^\dagger=(\delta\psi)^\dagger$, we get
$$
\eqalignno{\delta{\bf L}
&={\bf Tr}\,\left\{{I\over 2}~\left[\delta\psi^\dagger \hat
D_0\psi+\psi^\dagger\delta(\hat D_0\psi)-\delta(\hat
D_0\psi^\dagger)\psi-(\hat D_0\psi^\dagger)\delta\psi\right]\right.\cr
&~~~~~~\left.-m(\delta\psi^\dagger\psi
+\psi^\dagger\delta\psi)-\lambda_0I\delta B_0\right\}\cr
&={\bf Tr}\,\left(2(\delta\psi)^\dagger(I\hat
D_0\psi-m\psi)-I[\{\psi,\psi^\dagger\}_++\lambda_0]\delta
B_0\right)+{\partial\over\partial t}~{\bf Tr}\,
(I\psi^\dagger\delta\psi)~.&(25b)\cr}
$$
We have here explicitly used properties (ii) and (iii) of ${\bf Tr}$
discussed in the preceding section, with the appearance of the
anti--commutator $\{\psi ,\psi^\dagger\}_+$ a direct result of the effect
of the $(-1)^F$ factor on the reordering of fermion factors inside ${\bf
Tr}$.  The time derivative term in Eq.~(25b) makes no contribution
to the action ${\bf S}$, and so the Euler--Lagrange equations following
from $\delta{\bf S}=0$ can be read off directly from
Eq.~(25b),
giving the dynamical equation for $\psi$
$$
I\hat D_0\psi=m\psi~,
\eqno\rm (25c)
$$
together with the constraint
$$
\{\psi,\psi^\dagger\}_+=-\lambda_0~.
\eqno\rm (25d)
$$
Equation~(25d) is just the standard canonical anti--commutator for a
fermion degree of freedom when we identify $-\lambda_0=\hbar$ ($=1$ in
microscopic units).  We note that this determination of $\lambda_0$ has
the opposite sign from that found above in the bosonic example; we shall
say more about this shortly.  Finally, we remark that because the Hilbert
space for a single fermion degree of freedom consists of a zero fermion state
$|0\rangle\in V_\GH^+$ and a one fermion state $|1\rangle\in V_\GH^-$, the
effect of the $(-1)^F$ in the definition of ${\bf Tr}$ is to give
$$
{\bf Tr}\,1=1-1=0~,
\eqno\rm (25e)
$$
and so taking ${\bf Tr}$ of the left-- and right--hand sides of Eq.~(25d)
gives a consistent equation for non--zero $\lambda_0$,
$$
0={\bf Tr}\,(\psi\psi^\dagger-\psi\psi^\dagger)={\bf
Tr}\,\{\psi,\psi^\dagger\}_+
=-\lambda_0{\bf Tr}\,1=-\lambda_00~.
\eqno\rm (25f)
$$

{}From the time derivative term in Eq.~(25b), we identify the momentum
canonical to $\psi$ as
$$
p_\psi ={\delta{\bf L}\over\delta\dot\psi}=I\psi^\dagger~.
\eqno\rm (26a)
$$
Since by property (ii) of ${\bf Tr}$ we have
$$
{\bf Tr}(I\psi^\dagger\dot\psi)={\bf Tr}\left[\left({I\over
2}~\psi^\dagger\dot\psi\right)+\left({I\over
2}~\psi^\dagger\dot\psi\right)^\dagger\right]={\bf Tr}\left[{I\over
2}~(\psi^\dagger\dot\psi -\dot\psi^\dagger\psi)\right]~,
\eqno\rm (26b)
$$
the total trace Hamiltonian becomes
$$
\eqalignno{{\bf H}
&={\bf Tr}\,(I\psi^\dagger\dot\psi)-{\bf L}={\bf Tr}\,\left[{I\over
2}~(\psi^\dagger\dot\psi-\dot\psi^\dagger\psi)\right]-{\bf L}\cr
&={\bf Tr}\,\left(I[\{\psi,\psi^\dagger\}_++\lambda_0]\,B_0
+m\psi^\dagger\psi\right)~,&(26c)\cr}
$$
which simplifies in the Hamiltonian gauge $B_0=0$ to
$$
{\bf H}={\bf Tr}\,(m\psi^\dagger\psi)={\bf Tr}\,(-Im\,p_\psi\psi)~.
\eqno\rm (26d)
$$
Hence taking operator variations, we get
$$
\delta {\bf H}={\bf Tr}\,(-Im\,\delta p_\psi\psi-Im\,p_\psi\delta\psi)={\bf Tr}
\,[Im\,(\psi\delta p_\psi-p_\psi\delta\psi)]~,
\eqno\rm (26e)
$$
and so Eq.~(5f) becomes
$$
I\dot\psi^\dagger=\dot p_\psi=-{\delta {\bf
H}\over\delta\psi}
= Im\,p_\psi=-m\psi^\dagger~,~~~~~~
\dot\psi=-{\delta{\bf H}\over\delta p_\psi} =-Im\,\psi~,
\eqno\rm (26f)
$$
in agreement with Eq.~(25c) and its adjoint when $B_0=0$.  As in our
bosonic example, in the fermionic case there is also an operator
Hamiltonian
$$
H=m\psi^\dagger\psi
\eqno\rm (26g)
$$
which, with $\{\psi,\psi^\dagger\}_+=-\lambda_0$, obeys
$$
{1\over -I\lambda_0}~[\psi,H]=-Im\,\psi=\dot\psi~,~~~~~~
{1\over -I\lambda_0}~[\psi^\dagger,H]=Im\,\psi^\dagger=\dot\psi^\dagger~.
\eqno\rm (26h)
$$
Hence the dynamics of Eq.~(26f) is unitary, again permitting an operator
gauge transformation from the Heisenberg picture to the Schr\"odinger
picture.

We see, then, that the standard quantum mechanics of a single bosonic
and a single fermionic degree of freedom, including the canonical
commutator and anti--commutator, follows from the operator gauge
invariant total trace Lagrangian formalism.
Let us now examine what happens when more than one degree of freedom
is present.  Focusing only on the structure of the kinetic terms, for a
set $\{q_r\},\{\psi_s\}$ of bosonic and fermionic variables, we have the
conventional Lagrangian
$$
L^c_{\{q_r\},\{\psi_s\}}=\sum_{r=1}^R~{1\over 2}~\dot
q_r^2+\sum_{s=1}^S~{I\over
2}~(\psi_s^\dagger\dot\psi_s-\dot\psi_s^\dagger\psi_s)~.
\eqno\rm (27a)
$$
There is now more than one way to extend Eq.~(27a) into an operator
gauge invariant Lagrangian, depending on whether we require invariance
when all variables are subject to the same operator unitary
transformation, or are subject to independent operator unitary
transformations.  When all variables are subject to the same
transformation,
$$
q_r\to Uq_rU^\dagger~,~~~~~~\psi_s\to
U\psi_sU^\dagger~,~~~~~~UU^\dagger=U^\dagger U=1~,
\eqno\rm (27b)
$$
we achieve covariance of the Lagrangian by replacing
$L^c_{\{q_r\},\{\psi_s\}}$ by $L^{(1)}_{\{q_r\},\{\psi_s\}}$, with
$$
L^{(1)}_{\{q_r\},\{\psi_s\}}=\sum_{r=1}^R~{1\over 2}~(\hat
D_0q_r)^2+\sum_{s=1}^S~{I\over 2}~[\psi_s^\dagger\hat D_0\psi_s-(\hat
D_0\psi_s^\dagger)\psi_s]~,
\eqno\rm (27c)
$$
where $\hat D_0$ acts on the bosonic and fermionic degrees of freedom as
in Eqs.~(16c) and (24c).  The total trace Lagrangian and action
$$
{\bf L}^{(1)}_{\{q_r\},\{\psi_s\}}={\bf
Tr}\,L^{(1)}_{\{q_r\},\{\psi_s\}}~,~~~~~~
{\bf S}^{(1)}_{\{q_r\},\{\psi_s\}}=\int_{-\infty}^\infty~dt\,
{\bf L}^{(1)}_{\{q_r\},\{\psi_s\}}~,
\eqno\rm (27d)
$$
are then invariant under Eq.~(27b), and the constraint equation
arising from the variation with respect to $B_0$ of
$$
{\bf S}={\bf S}^{(1)}_{\{q_r\},\{\psi_s\}}-\lambda_0{\bf S}_{B_0}~,
\eqno\rm (27e)
$$
with ${\bf S}_{B_0}$ defined as in Eqs.~(18b) and (19a), is [cf
Eqs.~(20b) and (25b)]
$$
\sum_{r=1}^R~[q_r,\hat
D_0q_r]-I\sum_{s=1}^S~\{\psi_s,\psi_s^\dagger\}_+-\lambda_0I=0~.
\eqno\rm (27f)
$$
We see that the individual canonical commutators and anti--commutators
are not determined, but only the linear combination given by the sum of
bosonic commutators minus the sum of fermionic anti--commutators.  The
constraint of Eq.~(27f), as well as the dynamical equations of motion
for $\{q_r\}$ and $\{\psi_s\}$, are consistent with the imposition of
the canonical relations
$$
[q_r,\hat D_0q_r]=I~,~~~~~~\{\psi_s,\psi_s^\dagger\}_+=1~,
\eqno\rm (27g)
$$
for each $r$ and $s$, but do not require
these. In the language of the
conventional theory of constrained systems, Eqs.~(27g) are invariant
relations which are compatible with the Lagrangian constraints and
equations of motion.  We note the interesting fact that when the
numbers of bosonic and fermionic degrees of freedom are equal, as is the
case for a supersymmetric theory, and the canonical relations of
Eq.~(27g) are imposed, then the constraint of Eq.~(27f) is
satisfied with $\lambda_0=0$.

An alternative possibility is to require invariance of the total trace
Lagrangian under independent operator transformations of the canonical
variables,
$$
q_r\to U_rq_rU_r^\dagger~,~~~~~~\psi_s\to U_s\psi_sU_s^\dagger~,
$$
$$
U_rU_r^\dagger=U_r^\dagger U_r=U_sU_s^\dagger =U_s^\dagger U_s=1~.
\eqno\rm (28a)
$$
To achieve this, we introduce an independent covariant derivative for
each canonical variable,
$$
\dot q_r
\to \hat D_{0r}q_r\equiv {\partial\over\partial
t}~q_r+[B_{0r},q_r]~,~~~~~~\dot\psi_s\to \hat
D_{0s}\psi_s\equiv {\partial\over\partial t}~\psi_s+[B_{0s},\psi_s]~,
\eqno\rm (28b)
$$
with $B_{0r},~r=1,\ldots,R$ and $B_{0s},~s=1,\ldots,S$ independent
anti--self--adjoint operator gauge potentials.  We now replace the
Lagrangian $L^c_{\{q_r\},\{\psi_s\}}$ by
$$
L_{\{q_r\},\{\psi_s\}}^{(2)}=\sum_{r=1}^R~{1\over 2}~(\hat
D_{0r}q_r)^2+\sum_{s=1}^S~{I\over 2}~\left[\psi_s^\dagger\hat
D_{0s}\psi_s-(\hat D_{0s}\psi_s^\dagger)\psi_s\right]~,
\eqno\rm (28c)
$$
the individual terms of which transform covariantly (but now
independently) under Eq.~(28a),
$$
L_{\{q_r\},\{\psi_s\}}^{(2)}\to \sum_{r=1}^R~U_r{1\over 2}~(\hat
D_{0r}q_r)^2U_r^\dagger +\sum_{s=1}^S~U_s{I\over
2}~\left[\psi_s^\dagger\hat D_{0s}\psi_s-(\hat
D_{0s}\psi_s^\dagger)\psi_s\right]\,U_s^\dagger~.
\eqno\rm (28d)
$$
Equation~(28d) has the general form of Eq.~(8d), and
correspondingly, the total trace Lagrangian and action defined by
$$
{\bf L}_{\{q_r\},\{\psi_s\}}^{(2)}={\bf
Tr}\,L_{\{q_r\},\{\psi_s\}}^{(2)}~,~~~~~~S_{\{q_r\},\{\psi_s\}}^{(2)}
=\int_{-\infty}^\infty~dt {\bf L}_{\{q_r\},\{\psi_s\}}^{(2)}
\eqno\rm (28e)
$$
are invariant under Eq.~(28a).  For each gauge potential $B_{0r,s}$,
we can add an action term ${\bf S}_{B_{0r,s}}$ formed as in
Eqs.~(18b) and (19a), giving for the overall total trace action
$$
{\bf S}={\bf S}_{\{q_r\},\{\psi_s\}}^{(2)}-\sum_{r=1}^R~\lambda_{0r}
{\bf S}_{B_{0r}}-\sum_{s=1}^S~\lambda_{0s}{\bf S}_{B_{0s}}~.
\eqno\rm (28f)
$$
Varying with respect to each $B_{0r,s}$, we get the independent
constraints
$$
\eqalignno{[q_r,\hat D_{0r}q_r]-\lambda_{0r}I
&=0~,~~~~~~r=1,\ldots,R~,\cr
\{\psi_s,\psi_s^\dagger\}+\lambda_{0s}
&=0~,~~~~~~s=1,\ldots,S~, &(28g)\cr}
$$
which are closer in structure to the usual canonical commutators than
the single constraint of Eq.~(27f).  Of course, when we impose the
requirement of invariance under independent operator transformations as
in Eq.~(28a), the interaction terms in the total trace Lagrangian are
much more tightly restricted in form than when we impose only the global
invariance of Eq.~(27b).  In field theory applications, the
indices $r,s$ are typically composite indices, indicating both the
spatial coordinate value $\vec x$ and the particular field component at
$\vec x$.  In this case, the constraints associated
with $B_0$ have a structure intermediate in form between those of
Eqs.~(27f) and (28g):  the constraints at different values of
$\vec x$ are independent, but at each $\vec x$ consist of a sum
of contributions from the various bosonic and fermionic field components
present in the theory, evaluated at that value of $\vec x$.

Various further examples of operator gauge invariant quantum mechanical and
quantum field systems in
complex quantum mechanics, including models with independent left and right
gaugings as in Eq.~(9b), and operator gauge invariant extensions
of Yang--Mills theory, are given in [1].  In fact, we show there that all
of the basic field theory building blocks of the standard model can be
imbedded in operator gauge invariant theories; this raises the question of
studying operator gauge invariant extensions of the full standard model,
to see if useful insights (such as restrictions on the parameters, or
new calculational methods) can be obtained.
\bigskip
\def\boldcal#1{\hbox{\bfc #1}}
\font\bfc=cmbsy10 scaled\magstep1

\bigskip
\centerline{\bf 4. Operator gauge invariant quaternionic field theories}

We turn now to the construction of operator gauge invariant theories in
quaternionic quantum mechanics, which is distinguished from the complex
case discussed in the preceding section by the non--existence of a
left--acting $I$ which commutes with all operators.
For what follows only a few facts about quaternionic Hilbert space are
needed,
which we now proceed to state.
Quaternionic Hilbert space $V_\GH$ is closed under taking linear
combinations of ket vectors when right multiplied by quaternion scalars.  A
quaternion scalar is a number $q$ of the form
$$
\eqalignno{
& q=q_0+iq_1+jq_2+kq_3~,~~~~~~q_{0,1,2,3}~~{\rm real}~,\cr
& i^2=j^2=k^2=-1~,~~~~~~ij=-ji=k~,~~~~~~jk=-kj=i~,~~~~~~ki=-ik=j~,&(29a)\cr}
$$
from which one finds that quaternion multiplication, while non--commutative
[i.e., $q_1q_2\not= q_2q_1$ in general], is associative [i.e.,
$q_1(q_2q_3)=(q_1q_2)q_3$].  The conjugate $\bar q$ and modulus $|q|$ of
a quaternion are defined by
$$
\eqalignno{\bar q
&=q_0-iq_1-jq_2-kq_3~,\cr
|q| &=(\bar qq)^{1/2}=(q\bar q)^{1/2}=(q_0^2+q_1^2+q_2^2+q_3^2)^{1/2}~,
&(29b)\cr}
$$
and by Eq.~(29a), the conjugate of a product is the product of the conjugates
in reverse order,
$$
\overline{q_1q_2}=\bar q_2\bar q_1~.
\eqno\rm (29c)
$$
Defining the adjoint in $V_\GH$ as the quaternion conjugate of the matrix (or
operator) transpose, the adjoint of a ket vector is a bra vector, the adjoint
of a scalar is its conjugate, and from Eq.~(29c) one finds the usual rule
that the adjoint of any product of multiple factors is the product of the
adjoints of the factors in reverse order.  The inner product $\langle
f|g\rangle$
is quaternion--valued, and so any two inner products $\langle f|g\rangle$ and
$\langle f^\prime|g^\prime\rangle$ do not in general commute with one another
in quaternionic Hilbert space (whereas in complex quantum mechanics they do).
In other words, in quaternionic Hilbert space not only do operators not
commute with one another, but their {\it matrix elements} as well are
non--commutative.

For any complete orthonormal set of states $\{|n\rangle\}$, one defines
[12,13] the anti--self--adjoint left--acting algebra $I,J,K$ by
$$
(I,J,K)=\sum_n~|n\rangle\,(i,j,k)\,\langle
n|=-(I^\dagger,J^\dagger,K^\dagger)~,
\eqno\rm (29d)
$$
which obeys an isomorphic image of the algebra of right--acting quaternion
scalars,
$$
\eqalignno{I^2
&=J^2=K^2=-1=-\sum_n~|n\rangle\,\langle n|~,\cr
IJ &=-JI=K~,~~~~~~JK=-KJ=I,~~~~~~KI=-IK=J~. &(29e)\cr}
$$
For an arbitrary quaternion operator ${\cal O}$ which acts on kets from the
left
(all quaternion quantum fields discussed below fit this description), we can
introduce [12,13] a set of ``formally real'' components ${\cal O}_A,~
A=0,1,2,3$ defined by
$$
\eqalignno{{\cal O}_0
&={1\over 4}~({\cal O}-I{\cal O}I-J{\cal O}J-K{\cal O}K)~,\cr
{\cal O}_1 &=-{1\over 4}~(I{\cal O}+{\cal O}I-J{\cal O}K+K{\cal O}J)~,\cr
{\cal O}_2 &=-{1\over 4}~(J{\cal O}+{\cal O}J-K{\cal O}I+I{\cal O}K)~,\cr
{\cal O}_3 &=-{1\over 4}~(K{\cal O}+{\cal O}K-I{\cal O}J+J{\cal O}I)~,
&(29f)\cr}
$$
so--called because they obey
$$
{\cal O}={\cal O}_0+I{\cal O}_1+J{\cal O}_2+K{\cal O}_3~,~~~~~~
[{\cal O}_A,~(I,J,K)]=0~,~~~~~~A=0,1,2,3~,
\eqno\rm (29g)
$$
but they do not in general commute with one another.  Finally, we note that
$-{1\over 2}~I,-{1\over 2}~J,-{1\over 2}~K$ are a set of anti--self--adjoint
generators of a one--dimensional quaternionic irreducible representation [6,14]
of the group $SU(2)$; since the smallest non--trivial complex irreducible
representation of $SU(2)$ is the two--dimensional spinor representation, this
means that any quaternionic quantum field theory based on gauging the
one--dimensional quaternionic representation of $SU(2)$ is irreducibly
quaternionic, and cannot be reduced to a complex quantum field theory.  All
of the field equations discussed in this section have this character.

With these facts in mind, we are ready to begin the construction of
operator gauge invariant quaternionic field models.
We begin with the field theory of a quaternionic scalar field
$\phi$, which is not restricted to be self--adjoint (or
anti--self--adjoint), and which is subjected to independent left and right
local gaugings,\footnote{$^*$}{An historical
discussion of various gaugings which have been
proposed for quaternionic fields, and references, is given in
Sec.~11.2 of [1].
\smallskip
{}~~~~~Our metric convention is $g_{00}=-1,~g_{11}=g_{22}=g_{33}=1$.}
$$
\phi\to U\phi U^{\prime\dagger}~,~~~~~~UU^\dagger=U^\dagger
U=U^\prime U^{\prime\dagger}=U^{\prime\dagger}U^\prime =1~.
\eqno\rm (30a)
$$
Introducing anti--self--adjoint gauge potentials $B_\mu,B_\mu^\prime$
which transform as
$$
B_\mu\to UB_\mu U^\dagger -(\partial_\mu
U)U^\dagger~,~~~~~~B_\mu^\prime\to U^\prime B_\mu^\prime
U^{\prime\dagger}-(\partial_\mu U^\prime)U^{\prime\dagger}~,
\eqno\rm (30b)
$$
and the covariant derivative and field strengths
$$
\eqalignno{D_\mu\phi
&= \partial_\mu\phi+B_\mu\phi-\phi B_\mu^\prime~,\cr
F_{\mu\nu} &=\partial_\mu B_\nu-\partial_\nu B_\mu+[B_\mu,B_\nu]~,\cr
F_{\mu\nu}^\prime &=\partial_\mu B_\nu^\prime - \partial_\nu
B_\mu^\prime+[B_\mu^\prime,B_\nu^\prime]~, &(30c)\cr}
$$
which transform as
$$
D_\mu\phi\to U(D_\mu\phi) U^{\prime\dagger}~,~~~~~~F_{\mu\nu}\to
UF_{\mu\nu} U^\dagger~,~~~~~~F_{\mu\nu}^\prime\to U^\prime
F_{\mu\nu}^\prime U^{\prime\dagger}~,
\eqno\rm (30d)
$$
the operator gauge invariant total trace Lagrangian density for the
scalar field is
$$
\eqalign{\boldcal{L}
&=\boldcal{L}_\phi +\boldcal{L}_B+\boldcal{L}_{B^\prime}~,\cr
\boldcal{L}_\phi &={\bf Tr}\,\left\{{1\over
2}~[-(D_\mu\phi)^\dagger D^\mu\phi-m^2\phi^\dagger\phi]-{\lambda\over
4}~(\phi^\dagger\phi)^2\right\}~,\cr
\boldcal{L}_B &={\bf Tr}\,\left({1\over
4G^2}~F_{\nu\mu}F^{\nu\mu}\right)~,~~~~~~\boldcal{L}_{B^\prime}={\bf
Tr}\,\left({1\over
4(G^\prime)^2}~F_{\nu\mu}^\prime F^{\prime\nu\mu}\right)~.\cr}
\eqno\rm (30e)
$$
The total trace Lagrangian ${\bf L}$ and action ${\bf S}$ are formed
from $\boldcal{L}$ by the usual recipe
$$
{\bf L}=\int~d^3x\boldcal{L}~,~~~~~~{\bf S}=\int~dt\,{\bf L}~.
\eqno\rm (30f)
$$
Note that an action term analogous to ${\bf S}_{B_0}$ of Eq.~(19a) is
now not admissible, because the left--acting $I$ needed to construct
this term breaks the operator gauge invariance.  When we vary ${\bf S}$,
through $\delta F_{\mu\nu}$ and $\delta F_{\mu\nu}^\prime$ we encounter
the covariant derivatives $\hat D_\mu$ and $\hat D_\mu^\prime$,
$$
\eqalignno{\hat D_\mu{\cal O}
&=\partial_\mu{\cal O}+[B_\mu,{\cal O}]~,~~~~~~\hat D_\mu^\prime{\cal
O}=\partial_\mu{\cal O}+[B_\mu^\prime,{\cal
O}]~,\cr
\delta F_{\mu\nu} &=\hat D_\mu\delta B_\nu-\hat D_\nu\delta
B_\mu~,~~~~~~\delta F_{\mu\nu}^\prime =\hat D_\mu^\prime\delta
B_\nu^\prime-\hat D_\nu^\prime\delta B_\mu^\prime~, &(31a)\cr}
$$
and in integrating by parts we use the intertwining identities [1]
$$
\eqalignno{\hat D_\mu(\rho\eta^\dagger)
&= (D_\mu\rho)\eta^\dagger+\rho(D_\mu\eta)^\dagger~,\cr
\hat D_\mu^\prime(\rho^\dagger\eta) &=
(D_\mu\rho)^\dagger\eta+\rho^\dagger D_\mu\eta~,\cr
\partial_\mu {\bf Tr}\,(\rho\eta^\dagger) &={\bf
Tr}\,[(D_\mu\rho)\eta^\dagger+\rho(D_\mu\eta)^\dagger]~,\cr
\partial_\mu{\bf Tr}\,(\rho^\dagger\eta) &={\bf
Tr}\,[(D_\mu\rho)^\dagger\eta+\rho^\dagger D_\mu\eta]~, &(31b)\cr}
$$
with $\rho$ and $\eta$ either both bosonic or both fermionic in type.
Omitting further computational details, we get the operator equations of
motion
$$
\eqalignno{
{}~~~~~~~~~~~~~&D_\mu D^\mu\phi-(m^2+\lambda\phi\phi^\dagger)\phi=0~,\cr
&\hat D^\mu F_{\nu\mu}=G^2{\cal J}_\nu~,~~~~~~{\cal J}_\nu={1\over
2}~[\phi(D_\nu\phi)^\dagger-(D_\nu\phi)\phi^\dagger]~,\cr
&\hat D^{\prime\mu}F_{\nu\mu}^\prime=G^{\prime 2}{\cal
J}_\nu^\prime~,~~~~~~{\cal J}_\nu^\prime ={1\over 2}~[\phi^\dagger
D_\nu\phi-(D_\nu\phi)^\dagger\phi]~,&(32)\cr}
$$
in which the $\nu =0$ components of the gauge field equations are
constraints rather than dynamical equations.  Equations~(30e)
and (32) can be specialized to less general gaugings
of $\phi$.  For example, if we take $\phi$ to be gauged under
$$
\phi\to U\phi U^\dagger~,
\eqno\rm (33a)
$$
which is the most general allowed gauging when $\phi$ is self--adjoint,
the appropriate Lagrangian density is
$$
\boldcal{L}=\hat{\boldcal{L}}_\phi+\boldcal{L}_B~,~~~~~~
\hat{\boldcal{L}}_\phi={\bf Tr}\,\left\{{1\over 2}~[-(\hat
D_\mu\phi)^\dagger\hat D^\mu\phi-m^2\phi^\dagger\phi]-{\lambda\over
4}~(\phi^\dagger\phi)^2\right\}~,
\eqno\rm (33b)
$$
and the corresponding equations of motion are
$$
\eqalignno{
&~~~~~~~~\hat D_\mu\hat
D^\mu\phi-(m^2+\lambda\phi\phi^\dagger)\phi =0~,\cr
\hat D^\mu F_{\nu\mu} &={1\over 2}~G^2\left[\phi(\hat
D_\nu\phi)^\dagger-(\hat D_\nu\phi)\phi^\dagger+\phi^\dagger\hat
D_\nu\phi-(\hat D_\nu\phi)^\dagger\phi\right]~. &(33c)\cr}
$$

We turn next to the case of quaternionic fermion fields, working always in the
Majorana representation for the Dirac matrices, starting again
with the most general gauging, in which there are two fermions
$\psi_{(1)},\psi_{(2)}$ transforming under independent left and right local
gauge transformations as
$$
\psi_{(1)}\to U\psi_{(1)}U^{\prime\dagger}~,~~~~~~\psi_{(2)}\to
U\psi_{(2)} U^{\prime\dagger}~.
\eqno\rm (34a)
$$
The operator gauge invariant
total trace Lagrangian density is
$$
\boldcal{L}=\boldcal{L}_{\psi_{(1,2)}}+\boldcal{L}_B+
\boldcal{L}_{B^\prime}~,
\eqno\rm (34b)
$$
with $\boldcal{L}_B$ and $\boldcal{L}_{B^\prime}$ as in
Eq.~(30e), and with $\boldcal{L}_{\psi_{(1,2)}}$ given by
$$
\eqalignno{\boldcal{L}_{\psi_{(1,2)}}={\bf Tr}
&\left\{{1\over 2}~
\left[\psi_{(2)}^\dagger\gamma^0\gamma^\mu D_\mu\psi_{(1)}
+(D_\mu\psi_{(1)})^\dagger\gamma^0\gamma^\mu\psi_{(2)}-
\psi_{(1)}^\dagger\gamma^0\gamma^\mu
D_\mu\psi_{(2)}-(D_\mu\psi_{(2)})^\dagger\gamma^0\gamma^\mu\psi_{(1)}
\right]\right.\cr
&\left.~~~~~~+m\left(\psi_{(2)}^\dagger\,i\gamma^0\psi_{(1)}
-\psi_{(1)}^\dagger\,i\gamma^0\psi_{(2)}\right)\right\}~.&(34c)\cr}
$$
Varying the action ${\bf S}$ [still related to $\boldcal{L}$ by
Eq.~(30f)] and recalling that in the Majorana representation
$\gamma^0\gamma^\mu$ and $i\gamma^0$
are, respectively, real symmetric and real skew--symmetric matrices, we
get the operator equations of motion
$$
\eqalignno{
&~~(\gamma^0\gamma^\mu D_\mu+mi\gamma^0)\psi_{(1)}=0~,~~~~~~
(\gamma^0\gamma^\mu D_\mu+mi\gamma^0)\psi_{(2)}=0~,\cr
&\hat D^\mu F_{\nu\mu}=G^2{\cal J}_\nu~,~~~~~~{\cal
J}_\nu=\psi_{(1)}^T\gamma_\nu^T\gamma^{0T}\psi_{(2)}^{\dagger
T}-\psi_{(2)}^T\gamma_\nu^T\gamma^{0T}\psi_{(1)}^{\dagger T}~,\cr
& \hat D^{\prime\mu}F_{\nu\mu}^\prime= -(G^\prime)^2{\cal
J}_\nu^\prime~,~~~~~~{\cal J}_\nu^\prime=\psi_{(1)}^\dagger
\gamma^0\gamma_\nu\psi_{(2)}-\psi_{(2)}^\dagger\gamma^0\gamma_\nu\psi_{(1)}~,
&(35)\cr}
$$
with $T$ indicating Dirac index (but {\it not} operator) transposition.
Again, the $\nu=0$
components of the gauge field equations are constraints.  As in the
boson case, we can readily specialize the two fermion model to the less
general gauging
$$
\psi_{(1)}\to U\psi_{(1)} U^\dagger~,~~~~~~\psi_{(2)}\to U\psi_{(2)}
U^\dagger~,
\eqno\rm (36a)
$$
for which the appropriate Lagrangian density is
$$
\eqalignno{\boldcal{L}
&=\hat{\boldcal{L}}_{\psi_{(1,2)}}+\boldcal{L}_B~,\cr
\hat{\boldcal{L}}_{\psi_{(1,2)}}
&={\bf Tr}\left\{{1\over 2}~
\left[\psi_{(2)}^\dagger\gamma^0\gamma^\mu
\hat D_\mu\psi_{(1)}+(\hat D_\mu\psi_{(1)})^\dagger\gamma^0\gamma^\mu
\psi_{(2)}-\psi_{(1)}^\dagger\gamma^0\gamma^\mu
\hat D_\mu\psi_{(2)}-(\hat D_\mu\psi_{(2)})^\dagger\gamma^0\gamma^\mu\psi_{(1)}
\right]\right.\cr
&\biggl.~~~~~~~~+m\left(\psi_{(2)}^\dagger\,i\gamma^0\psi_{(1)}
-\psi_{(1)}^\dagger\,i\gamma^0\psi_{(2)}\right)\biggr\}~,&(36b)\cr}
$$
and the corresponding equations of motion are
$$
\eqalign{
&(\gamma^0\gamma^\mu \hat D_\mu+mi\gamma^0)\psi_{(1)}=0~,~~~~~~
(\gamma^0\gamma^\mu \hat D_\mu+mi\gamma^0)\psi_{(2)}=0~,\cr
\hat D^\mu
F_{\nu\mu}&=G^2\left[\psi_{(1)}^T\gamma_\nu^T\gamma^{0T}\psi_{(2)}^{\dagger
T}-\psi_{(2)}^T\gamma_\nu^T\gamma^{0T}\psi_{(1)}^{\dagger
T}-\psi_{(1)}^\dagger\gamma^0\gamma_\nu\psi_{(2)}+\psi_{(2)}^\dagger
\gamma^0\gamma_\nu\psi_{(1)}\right]~.\cr}
\eqno\rm (36c)
$$

We can also form fermionic models with a single fermion field, in which
either the
left or the right gauge invariance is restricted to be a complex gauge
invariance.  With the right gauge invariance restricted, we get
$$
\boldcal{L}_\psi={\bf Tr}\,\left\{{I^\prime\over
2}~\left[\psi^\dagger\
\gamma^0\gamma^\mu
D_\mu\psi-(D_\mu\psi)^\dagger\gamma^0\gamma^\mu\psi\right]+I^\prime m
\psi^\dagger i\gamma^0\psi\right\}~,
\eqno\rm (37a)
$$
with $I^\prime$ a space--time independent left algebra operator, and
with $U^\prime$ and $B_\mu^\prime$ restricted to be $\GC(1,I^\prime)$.
Similarly, with the left gauge invariance restricted, we have
$$
\boldcal{L}_\psi^\prime ={\bf Tr}\,\left\{{1\over
2}\left[\psi^\dagger I\gamma^0\gamma^\mu
D_\mu\psi-(D_\mu\psi)^\dagger\gamma^0\gamma^\mu
I\psi\right]+m\psi^\dagger I\,i\gamma^0\psi\right\}~,
\eqno\rm (37b)
$$
with $I$ a space--time independent left algebra operator, and with $U$
and $B_\mu$ restricted to be $\GC(1,I)$.  When $B_\mu$ is so restricted,
an action term ${\bf S}_{B_\mu}$ analogous to
Eq.~(19a) can be included in
the total action, and similarly for $B_\mu^\prime$ in the case of
Eq.~(37a).

Before proceeding to the total trace Hamiltonian form of the dynamics,
we discuss a number of issues which can be addressed directly from the
total trace Lagrangian and the equations of motion.
\item{(1)}
We begin by contrasting the quaternionic gauge field structure with that
of a conventional Yang--Mills gauge field.  Let $1,E_A,~A=1,2,3$ be a
space--time--independent left algebra basis, and let us use Eqs.~(29f,g)
to expand the gauge potential $B_\mu$ and the corresponding
field--strength $F_{\mu\nu}$ over this basis,
$$
B_\mu=B_{0\mu}+\sum_{A=1}^3~B_{A\mu}E_A~,~~~~~~F_{\mu\nu}
=F_{0\mu\nu}+\sum_{A=1}^3~F_{A\mu\nu}E_A~,
\eqno\rm (38a)
$$
with the expansion coefficients
$B_{0\mu},B_{A\mu},F_{0\mu\nu},F_{A\mu\nu}$ formally real,
$$
[B_{0\mu},E_C]=[B_{A\mu},E_C]=[F_{0\mu\nu},E_C]=[F_{A\mu\nu},E_C]=0~.
\eqno\rm (38b)
$$
We recall, however, from the discussion beginning the section,
that in general
the expansion coefficients $B_{0\mu},B_{A\mu},\ldots$ are still
operators which do not commute with one another.  Substituting
Eqs.~(38a) into the formula of Eq.~(30c) which relates the gauge
field strength $F_{\mu\nu}$ to the gauge potential $B_\mu$, we find that
the expansion coefficients $F_{A\mu\nu},A=0,\ldots,3$ are related to
the $B_{A\mu}, A=0,\ldots,3$ by
$$
\eqalign{F_{0\mu\nu}
&=\partial_\mu B_{0\nu}-\partial_\nu
B_{0\mu}+[B_{0\mu},B_{0\nu}]-\sum_{A=1}^3~[B_{A\mu},B_{A\nu}]~,\cr
F_{A\mu\nu} &=\partial_\mu B_{A\nu} -\partial_\nu
B_{A\mu}+[B_{A\mu},B_{0\nu}]-[B_{A\nu},B_{0\mu}]+\sum_{B,C=1}^3~
\varepsilon_{ABC}\{B_{B\mu},B_{C\nu}\}_+~.~~~~~~ \cr}
\eqno\rm (38c)
$$
If $B_{A\mu},A=0,\ldots,3$ all commute with one another, Eq.~(38c)
would reduce to $U(1)$ and $SU(2)$ conventional gauge field structures,
$$
\eqalignno{F_{0\mu}
&=\partial_\mu B_{0\nu}-\partial_\nu B_{0\mu}~,\cr
F_{A\mu\nu} &=\partial_\mu B_{A\nu}-\partial_\nu
B_{A\mu}+2\sum_{B,C=1}^3~\varepsilon_{ABC}B_{B\mu}B_{C\nu}~.&(38d)\cr}
$$
But in the general case with non--commuting formally real components
$B_{A\mu}$, Eqs.~(38c) are not equivalent to
Eqs.~(38d).  Equations~(38c) represent only part of
the complete system of equations following from the total trace
Lagrangians of Eqs.~(30e), (34b,c), etc.  It is straightforward
to reexpress all of the remaining field equations in terms of formally
real components with respect to the left--acting algebra $1,E_A$.
\item{}
{}~~~~~Because the $B_{A\mu}$ are quaternionic operators, they can
themselves be expanded over formally real components with respect to a
second left--acting algebra $1,E_B^{(1)}$ which commutes with $1,E_A$,
$$
\eqalignno{B_{A\mu}
&=B_{0A\mu}+\sum_{B=1}^3~B_{BA\mu}E_B^{(1)}~,\cr
[B_{BA\mu},E_C]&=[B_{BA\mu},E_C^{(1)}]=0~,\cr
[E_B,E_C^{(1)}]&=0~,~~~~~~A,B,C=0,1,2,3~. &(39a)\cr}
$$
This process can be continued to any order, giving in $n$th order a
multi--quaternion expansion of the form (with
$E_0=E_0^{(1)}=\ldots=E_0^{(n)}=1$)
$$
B_\mu=\sum_{A=0}^3~\sum_{A_1=0}^3~\ldots\sum_{A_n=0}^3~B_{A_n\ldots
A_1A\mu}E_AE_{A_1}^{(1)}\ldots E_{A_n}^{(n)}~,
\eqno\rm (39b)
$$
with, for all index values, vanishing commutators
$$
[E_B,E_C^{(r)}]=[E_B^{(r)},E_C^{(s)}]=[B_{A_n\ldots A_1A\mu},E_B]=
[B_{A_n\ldots A_1A\mu},E_B^{(r)}]=0~.
\eqno\rm (39c)
$$
Note that the occurrence of left acting multi--quaternion algebras does
not imply that probability amplitudes belong to a non--division algebra.
 The algebra of right--acting scalars remains just $1,i,j,k$, with the
possibility of multiple commuting left--acting algebras arising because
these are quaternionic Hilbert space operators, rather than scalars.
For example, in as small as a four dimensional quaternionic Hilbert
space there are two mutually commuting left algebra bases,
$$
\eqalign{1
&=\left[\matrix{1&0&0&0\cr
0&1&0&0\cr
0&0&1&0\cr
0&0&0&1\cr}\right]~,E_1=i1,~E_2=j1,~E_3=k1~,\cr
E_1^{(1)} &=\left[\matrix{0&0&0&1\cr
0&0&-1&0\cr
0&1&0&0\cr
-1&0&0&0\cr}\right]~,
E_2^{(1)}=\left[\matrix{0&0&1&0\cr
0&0&0&1\cr
-1&0&0&0\cr
0&-1&0&0\cr}\right]~,
E_3^{(1)}=\left[\matrix{0&-1&0&0\cr
1&0&0&0\cr
0&0&0&1\cr
0&0&-1&0\cr}\right]~.\cr}
\eqno\rm (39d)
$$
Properties of multi--quaternion algebras have been studied in a series of
papers by Razon and Horwitz [15--17], and by various authors [18--21]
who classified the
$U(2)$ case of the ``color charge algebras'' introduced by Adler [22].
(The color charge algebras
correspond to multi--quaternion bases $E_{A_1}^{(1)}\ldots
E_{A_n}^{(n)}$, in which the indices $A_1,\ldots, A_n$ are contracted so
as to leave a single free index $A$.)
The latter calculations indicate that large Lie algebras are
readily built up from multi--quaternion bases.
\item{(2)}  In the fermionic Lagrangians of Eqs.~(34c) and
(36b), the gauge bosons couple to vector currents only, and so there
are no chiral anomalies and also no Witten [23] anomalies. Since
the $\gamma_5$ matrix in Majorana representation is
imaginary, attempting to split Eqs.~(34c) and (36b) into
chiral components would break the operator gauge invariance.  In other
words, insisting on a maximal operator gauge invariance in quaternionic
field theory excludes chiral fermions.
\item{}
{}~~~~~At the same time, since $i\gamma_5$ is real and
anti--self--adjoint, when the mass $m$ is zero the fermion total trace
Lagrangians written above are invariant under
$$
\eqalignno{\psi_{(1,2)}
&\to e^{i\gamma_5\beta}\psi_{(1,2)}~,~~~~~\psi_{(1,2)}^\dagger\to
\psi_{(1,2)}^\dagger e^{-i\gamma_5\beta}~,\cr
\psi &\to e^{i\gamma_5\beta}\psi~,~~~~~~~~~\psi^\dagger\to \psi^\dagger
e^{-i\gamma_5\beta}~, &(39e)\cr}
$$
with $\beta$ a real $c$--number which is independent of $x$.  Thus, when
$m=0$, the fermion models all have a chiral
symmetry.
\item{(3)}
Let us next address the issues of the spin--statistics connection and
discrete symmetries in quaternionic field theory.
On spin and statistics, we have little to
say, beyond the fact that the conventional spin--statistics connection
has been assumed in including the $(-1)^F$ factor in the definition of
${\bf Tr}$.  We have not made a study of the possibility of an abnormal
spin--statistics connection, but note that this would lead to the usual
pathologies upon specializing a quaternionic field theory back to a
complex one.
\item{}
{}~~~~~To study discrete symmetries of the Lagrangians written down above,
we use some standard Dirac matrix apparatus.  We begin
with parity $(P)$, and readily find that under the substitutions
$$
\eqalignno{B_0(\vec x,x^0)
&\to B_0(-\vec x,x^0)~,~~~~~~B_0^\prime(\vec x,x^0)\to B_0^\prime(-\vec
x,x^0)~,\cr
B_\ell(\vec x,x^0) &\to -B_\ell(-\vec x,x^0)~,~~~~~~B_\ell^\prime(\vec
x,x^0)\to -B_\ell^\prime(-\vec x,x^0)~,\cr
\phi(\vec x,x^0)&\to \eta_P\phi(-\vec x,x^0)~,\cr
\psi_{(1,2)}(\vec x,x^0)&\to \eta_P^\prime i\gamma^0\psi_{(1,2)}(-\vec
x,x^0)~,~~~~~~
\psi(\vec x,x^0)\to \eta_P^{\prime\prime}i\gamma^0\psi(-\vec
x,x^0)~,~~~~~~
&(40a)\cr}
$$
with $\eta_P,\eta_P^\prime$, and $\eta_P^{\prime\prime}$ arbitrary real
$c$--number phases, all of the total trace Lagrangian densities of this
section transform as
$$
\boldcal{L}(\vec x,x^0)\to \boldcal{L}(-\vec x,x^0)~,
\eqno\rm (40b)
$$
and the corresponding total trace actions are invariant.  Turning next
to time reversal $(T)$, the relevant substitutions are now
$$
\eqalignno{B_0(\vec x,x^0)
&\to -B_0(\vec x,-x^0)~,~~~~~~B_0^\prime(\vec x,x^0)\to
-B_0^\prime(\vec x,-x^0)~,\cr
B_\ell(\vec x,x^0)&\to B_\ell(\vec x,-x^0),~~~~~~B_\ell^\prime(\vec
x,x^0)\to B_\ell^\prime(\vec x,-x^0)~,\cr
\phi(\vec x,x^0) &\to \eta_T\phi(\vec x,-x^0)~,\cr
\psi_{(1)}(\vec x,x^0)&\to \eta_T^\prime A\psi_{(1)}(\vec x,-x^0)~,\cr
\psi_{(2)}(\vec x,x^0) &\to -\eta_T^\prime A\psi_{(2)}(\vec x,-x^0)~,\cr
\psi(\vec x,x^0) &\to \cases{\eta_T^{\prime\prime}A\psi(\vec
x,-x^0)J^\prime ~{\rm for}~\boldcal{L}_\psi\cr
\eta_T^{\prime\prime} JA\psi(\vec x,-x^0) ~{\rm for}~\boldcal{L}_\psi^\prime~,
\cr}~,&(41a)\cr}
$$
with $\eta_T,\eta_T^\prime,\eta_T^{\prime\prime}$ arbitrary real
$c$--number phases, with $A$ the Majorana representation time reversal
matrix, which is real, and with $J^\prime,J$ anti--commuting respectively with
$I^\prime, I$.  Under these substitutions all of the total trace
Lagrangian densities of this section transform as
$$
\boldcal{L}(\vec x,x^0)\to \boldcal{L}(\vec x,-x^0)~,
\eqno\rm (41b)
$$
and the corresponding total trace actions are again invariant.
\item{}
{}~~~~~Turning finally to charge conjugation $(C)$, we consider the
substitutions
$$
\eqalignno{B_\mu
&\to B_\mu^\prime~,~~~~~~B_\mu^\prime\to B_\mu~,\cr
\phi &\to \eta_C\phi^\dagger~,\cr
\psi_{(1)} &\to \eta_C^\prime\psi_{(1)}^{\dagger T}~,~~~~~~
\psi_{(2)} \to -\eta_C^\prime\psi_{(2)}^{\dagger T}~,\cr
\psi &\to \eta_C^{\prime\prime}\psi^{\dagger T}~,
&(42a)\cr}
$$
under which the covariant derivatives $D_\mu\phi, D_\mu\psi_{(1,2)}$,
and $D_\mu\psi$ transform as
$$
D_\mu\phi\to
\eta_C(D_\mu\phi)^\dagger~,~~~~~~D_\mu\psi_{(1,2)}\to(+,-)
\eta_C^\prime(D_\mu\psi_{(1,2)})^{\dagger T}~,~~~~~~
D_\mu\psi\to\eta_C^{\prime\prime}(D_\mu\psi)^{\dagger T}~,
\eqno\rm (42b)
$$
with $\eta_C, \eta_C^\prime$, and $\eta_C^{\prime\prime}$ arbitrary
real $c$--number phases and with $T$ the Dirac index transpose.  When we
impose a condition of equality on the gauge field couplings,
$$
G=G^\prime~,
\eqno\rm (42c)
$$
all of the total trace Lagrangian densities of this section are
invariant under the substitutions of Eq.~(42a),
$$
\boldcal{L} \to \boldcal{L}~.
\eqno\rm (42d)
$$
Since the gauge potentials $B_\mu$ and $B_\mu^\prime$ are interchanged
by the substitutions of Eq.~(42a), they are not $C$ eigenstates.  We
see, then, that when the requirement of $C$ invariance is imposed, the
models with independent left and right gaugings are left with a single
coupling constant $G$.\footnote{$^*$}{When $G\not= G^\prime$, the models do not
have a
$C$ or a $CPT$ symmetry; this does not contradict the usual $CPT$ theorem
because we do not make the locality assumption that the fields commute at
spacelike separations.  In [1] it is shown that when the Lagrangians
of this section are reinterpreted as complex field theory Lagrangians
through a decomposition of the fields into symplectic components,
then one finds a $C$ and a $CPT$ symmetry even for $G\not= G^\prime$, in
agreement with the usual $CPT$ theorem.}
\item{(4)}
By using the fact that $i\gamma_5$ is
real and anti--self--adjoint, we can construct a model with
self--adjoint Lagrangian using a single fermion field, without breaking
the bi--unitary operator gauge invariance.  The Lagrangian density for
this model is
$$
\boldcal{L}_\psi^5 ={\bf Tr}\,\left\{{1\over
2}~\psi^\dagger\gamma^0\gamma^\mu(i\gamma_5)D_\mu\psi
-(D_\mu\psi)^\dagger\gamma^0\gamma^\mu(i\gamma_5)\psi\right\}~;
\eqno\rm (42e)
$$
we do not include a mass term because, since $i\gamma_5$
anti--commutes with $i\gamma^0$, the expression
$$
\psi^\dagger i\gamma^0 i\gamma_5\psi
\eqno\rm (42f)
$$
is anti--self--adjoint, and vanishes inside ${\bf Tr}$.  It is easy to
check that under the transformations
$$
\eqalignno{P:~~~~~~\psi(\vec x,x^0)
&\to \eta_P i\gamma^0\psi(-\vec x,x^0)~,\cr
C:~~~~~~~~~~~~~~~\psi
&\to \eta_C\psi^{\dagger T}~,~~~~~~~~~~~~~~~~~~~\cr
T:~~~~~~\psi(\vec x,x^0)&\to \eta_T A\psi(\vec x,-x^0)~,
&(42g)\cr}
$$
together with the gluon sector transformations discussed above, the
Lagrangian density $\boldcal{L}_\psi^5$ is $P$ odd, $C$ odd, and $T$
even.  Although naively leading to conserved source currents for the
gauge gluons, the model $\boldcal{L}_\psi^5$ has chiral anomalies in
the usual complex canonical quantization, which suggests that it may
also be inconsistent in the more general total trace Lagrangian
dynamics.  This question requires further study.
\item{(5)}
Up to this point, our entire discussion has dealt with field theories in
flat space--time.  Since the total trace Lagrangians of this section are
all Lorentz invariant, they can be generalized to curved space--time by
the standard prescription of replacing the Minkowski metric
by a general metric $g_{\mu\nu}$, ordinary derivatives
$\partial_\lambda$ by covariant derivatives $\bigtriangledown_\lambda$
which commute with $g_{\mu\nu}$, etc.  When this is done, the source
term for the gravitational field equation will be a total trace
energy--momentum tensor ${\bf T}^{\mu\nu}$, defined by computing the
variation of the total trace action ${\bf S}$ under an infinitesimal
metric variation $g_{\mu\nu}\to g_{\mu\nu}+\delta g_{\mu\nu}$, according
to
$$
{\bf S}\to{\bf S}+\delta{\bf S}~,~~~~~~{\bf \delta}{\bf S}={1\over
2}~\int~d^4x[g(x)]^{1/2}{\bf T}^{\mu\nu}(x)\delta g_{\mu\nu}(x)~.
\eqno\rm (43a)
$$
(Here $g(x)$ is the negative of the determinant of the metric, and our
notation follows Weinberg [24].)  Standard arguments then
show that ${\bf T}^{\mu\nu}$ is covariantly conserved,
$$
\bigtriangledown_\mu{\bf T}^{\mu\nu}=0~,
\eqno\rm (43b)
$$
and, in the flat space--time limit, the spatial integrals of the
various components of ${\bf T}^{\mu\nu}$ give the total
trace Poincar\'e generators, for example,
$$
{\bf H}=\int~d^3x\,{\bf T}^{00}~.
\eqno\rm (43c)
$$
We conclude that quaternionic field theories described by total trace
actions can be consistently coupled to classical gravitation, but the
total trace structure of the gravitational source term differs from that
customarily assumed in the semi--classical theory of
gravitation.
\item{(6)}
It is interesting to ask whether the Lagrangian ${\bf L}$ of
Eqs.~(30f) and (34b,c) has fermionic symmetries, constructed in analogy with
the
fermionic symmetry of supersymmetric Yang--Mills theory [25,26].
This question can be
investigated by studying the change of ${\bf L}$ under field variations
parameterized by fermionic parameters, leading to a lengthy calculation,
the results of which are compactly summarized as covariant divergence
equations for the corresponding Noether currents.  Let
${\cal S}_{(1,2)}^\lambda$ be the fermionic currents
$$
{\cal S}_{(1,2)}^\lambda={1\over
2}~\left[\psi_{(1,2)}^\dagger\,{F^{\mu\nu}\over
G^2}+{F^{\prime\mu\nu}\over G^{\prime
2}}~\psi_{(1,2)}^\dagger\right]\gamma^0\gamma^\lambda {1\over
2}~[\gamma_\mu,\gamma_\nu]~,
\eqno\rm (44a)
$$
which transform under operator gauge transformations as
$$
{\cal S}_{(1,2)}^\lambda\to U^\prime {\cal S}_{(1,2)}^\lambda
U^\dagger~.
\eqno\rm (44b)
$$
Let us define the covariant derivative $\bar D_\lambda$ acting on a
general operator ${\cal O}$ as
$$
\bar D_\lambda{\cal O}=\partial_\lambda {\cal O}+B_\lambda^\prime {\cal
O}-{\cal O}B_\lambda~,
\eqno\rm (44c)
$$
so that when ${\cal O}$ gauge transforms as
$$
{\cal O}\to U^\prime {\cal O} U^\dagger~,
\eqno\rm (44d)
$$
$D_\lambda{\cal O}$ transforms covariantly as
$$
D_\lambda{\cal O}\to U^\prime (D_\lambda{\cal O})U^\dagger~.
\eqno\rm (44e)
$$
Comparing Eq.~(44c) with the definition of $D_\mu$ in Eq.~(30c),
we evidently have
$$
(D_\lambda {\cal O})^\dagger =\bar D_\lambda {\cal O}^\dagger~.
\eqno\rm (44f)
$$
Forming the covariant divergence $\bar D_\lambda {\cal
S}_{(1,2)}^\lambda$ and using Eq.~(44f), we get the identity
$$
\eqalignno{\bar D_\lambda
&{\cal S}_{(1,2)}^\lambda
={1\over 2}~\left[(D_\lambda \psi_{(1,2)})^\dagger {F^{\mu\nu}\over
G^2}+{F^{\prime\mu\nu}\over G^{\prime
2}}~(D_\lambda\psi_{(1,2)}^\dagger)\right]\gamma^0\gamma^\lambda {1\over
2}~[\gamma_\mu,\gamma_\nu]\cr
&+{1\over 2}~\left[\psi_{(1,2)}^\dagger{(\hat D_\lambda F^{\mu\nu})\over
G^2}+{(\hat D_\lambda^\prime F^{\prime\mu\nu})\over G^{\prime 2}}~
\psi_{(1,2)}^\dagger\right]\gamma^0\gamma^\lambda{1\over
2}~[\gamma_\mu,\gamma_\nu]~. &(45a)\cr}
$$
The first line of Eq.~(45a) can be simplified by using the Dirac
equations of Eq.~(35), expressed in the form
$$
(D_\lambda \psi_{(1,2)})^\dagger \gamma^0\gamma^\lambda = \psi_{(1,2)}^\dagger
mi\gamma^0~,
\eqno\rm (45b)
$$
to give
$$
{1\over 2}~\left[\psi_{(1,2)}^\dagger{F^{\mu\nu}\over
G^2}+{F^{\prime\mu\nu}\over G^{\prime 2}}~\psi_{(1,2)}^\dagger\right]\,
mi\gamma^0{1\over 2}~[\gamma_\mu,\gamma_\nu]~.
\eqno\rm (45c)
$$
The second line of Eq.~(45a) can be rearranged by substituting the
identity
$$
\gamma^\lambda{1\over
2}~[\gamma_\mu,\gamma_\nu]=\gamma_\mu\delta_\nu^\lambda-\gamma_\nu\delta_\mu^\lambda
- \varepsilon_{\mu\nu}^{~~\lambda\alpha}\gamma_\alpha i\gamma^5~.
\eqno\rm (45d)
$$
The contribution from $\varepsilon_{\mu\nu}^{~~\lambda\alpha}$ vanishes
by virtue of the Bianchi identities
$$
\hat D_\lambda F_{\mu\nu}+\hat D_\nu F_{\lambda\mu}+\hat
D_\mu F_{\nu\lambda}=0~,~~~~~~
\hat D_\lambda^\prime F_{\mu\nu}^\prime +\hat D_\nu^\prime
F_{\lambda\mu}^\prime + \hat D_\mu^\prime F_{\nu\lambda}^\prime =0~.
\eqno\rm (45e)
$$
The contribution from the Kronecker delta terms is
$$
\left[\psi_{(1,2)}^\dagger {\hat D_\nu F^{\mu\nu}\over G^2}+
{\hat D_\nu^\prime F^{\prime\mu\nu}\over G^{\prime
2}}~\psi_{(1,2)}^\dagger\right]\gamma^0\gamma_\mu~,
\eqno\rm (45f)
$$
which can be simplified using the gauge field equations of Eq.~(35)
to give
$$
\left[\psi_{(1,2)}^\dagger {\cal J}^\mu-{\cal J}^{\prime\mu}
\psi_{(1,2)}^\dagger\right]\gamma^0\gamma_\mu~.
\eqno\rm (45g)
$$
Thus, putting everything together, we have
$$
\eqalignno{\bar D_\lambda
&{\cal S}_{(1,2)}^\lambda ={1\over 2}\left(\psi_{(1,2)}^\dagger
{F^{\mu\nu}\over G^2}+{F^{\prime\mu\nu}\over G^{\prime 2}}~\psi_{(1,2)}^\dagger
\right)\,mi\gamma^0{1\over 2}~[\gamma_\mu,\gamma_\nu]\cr
&+ \left(\psi_{(1,2)}^\dagger{\cal J}^\mu -{\cal J}^{\prime\mu}
\psi_{(1,2)}^\dagger\right)\gamma^0\gamma_\mu~, &(46a)\cr}
$$
and we see that even when the fermion mass $m$ vanishes, the fermionic
currents ${\cal S}_{(1,2)}^\lambda$ are not covariantly conserved.
Suppose, however, that there is either an operator gauge, or an
asymptotic limit, in which the fermion fields $\psi_{(1,2)}$ have the
standard canonical anti--commutators of complex fields.  In such a
situation, we see from Eq.~(35) that we would have ${\cal J}^\mu =
{\cal J}^{\prime\mu}$, and the second line of Eq.~(46a) would reduce
to the singular commutator
$$
[\psi_{(1,2)}^\dagger,{\cal J}^\mu]\gamma^0\gamma_\mu~,
\eqno\rm (46b)
$$
which vanishes in dimensional
regularization.  This argument suggests that Eq.~(46a), despite the
presence of the gluon source current terms, may nonetheless have useful
content.

Let us now return to our main theme of total trace operator dynamics,
and construct the total trace Hamiltonian form of the dynamics following
from the scalar field Lagrangian of Eqs.~(30e,f) and the fermion field
Lagrangian of Eqs.~(34b,c).  From Eq.~(30e), we get
$$
\eqalignno{p_\phi
&={\delta{\bf L}\over\delta\dot\phi}=(D_0\phi)^\dagger~,\cr
p_{B_\ell} &={\delta{\bf L}\over\delta\dot B_\ell}=-{1\over
G^2}~F_{0\ell}~,~~~~~~p_{B^\prime_\ell} ={\delta {\bf L}
\over\delta\dot B_\ell^\prime}=-{1\over G^{\prime
2}}~F_{0\ell}^\prime~, &(47a)\cr}
$$
and so the total trace Hamiltonian density becomes
$$
\boldcal{H}={\bf Tr}\left[(D_0\phi)^\dagger\dot\phi-{1\over
G^2}~\sum_{\ell=1}^3~F_{0\ell}\dot B_\ell-{1\over G^{\prime
2}}~\sum_{\ell=1}^3~F_{0\ell}^\prime\dot
B_\ell^\prime\right]-\boldcal{L}~.
\eqno\rm (47b)
$$
Substituting
$$
\eqalign{\dot\phi
&=D_0\phi-B_0\phi+\phi B_0^\prime~,\cr
\dot B_\ell
&=F_{0\ell}+\hat D_\ell B_0~,~~~~~~\dot
B_\ell^\prime=F_{0\ell}^\prime+\hat D_\ell^\prime B_0^\prime~,\cr}
\eqno\rm (47c)
$$
forming the total trace Hamiltonian ${\bf H}$ and doing a spatial
integration by parts, we get
$$
{\bf H}=\int~d^3x\boldcal{H}={\bf H}_\phi+{\bf
H}_{B,{\cal J}_0}+{\bf H}_{B^\prime,{\cal J}_0^\prime}~,
\eqno\rm (48a)
$$
with
$$
\eqalign{{\bf H}_\phi
&=\int~d^3x{\bf Tr}\left[{1\over 2}~p_\phi^\dagger p_\phi+{1\over
2}~\sum_{\ell=1}^3~(D_\ell\phi)^\dagger (D_\ell\phi)+{1\over
2}~m^2\phi^\dagger\phi+{\lambda\over 4}~(\phi^\dagger\phi)^2\right]~,\cr
{\bf H}_{B,{\cal J}_0}
&=\int~d^3x{\bf Tr}\left[-{G^2\over
2}~\sum_{\ell=1}^3~(p_{B_\ell})^2-{1\over
2G^2}~\sum_{{\ell,m=1\atop\ell<m}}^3(F_{\ell m})^2-B_0 ({\cal
J}_0+\sum_{\ell=1}^3~\hat D_\ell p_{B_\ell})\right]~,\cr
{\bf H}_{B^\prime,{\cal J}_0^\prime}
&=\int~d^3x{\bf Tr}\left[-{G^{\prime 2}\over
2}~\sum_{\ell=1}^3~(p_{B_\ell^\prime})^2-{1\over 2G^{\prime
2}}\sum_{{\ell,m=1\atop\ell<m}}^3~(F_{\ell m}^\prime)^2-B_0^\prime({\cal
J}_0^\prime+\sum_{\ell=1}^3~\hat D_\ell^\prime
p_{B_\ell^\prime})\right]~,~~~~~\cr}
\eqno\rm (48b)
$$
and with ${\cal J}_0,{\cal J}_0^\prime$ the $0$ components of the boson
source currents ${\cal J}_\nu,{\cal J}_\nu^\prime$ given in
Eq.~(32).  Proceeding similarly in the fermion case, from
Eqs.~(34b,c) we get
$$
p_{\psi_{(1)}}={\delta{\bf L}\over\delta\dot\psi_{(1)}}=\psi_{(2)}^\dagger~,
{}~~~~~~p_{\psi_{(2)}}={\delta{\bf L}\over\delta\dot\psi_{(2)}}
=-\psi_{(1)}^\dagger~,
\eqno\rm (49a)
$$
and so the total trace Hamiltonian density becomes
$$
\boldcal{H}={\bf Tr}\left[ \psi_{(2)}^\dagger \dot\psi_{(1)}-
\psi_{(1)}^\dagger \dot\psi_{(2)}-{1\over G^2}\sum_{\ell=1}^3~F_{0\ell}
\dot B_\ell-{1\over G^{\prime 2}}\sum_{\ell=1}^3~F_{0\ell}^\prime \dot
B_\ell^\prime\right]-\boldcal{L}~.
\eqno\rm (49b)
$$
Substituting
$$
\dot\psi_{(1,2)}=D_0\psi_{(1,2)}-B_0\psi_{(1,2)}+\psi_{(1,2)}B_0^\prime~,
\eqno\rm (49c)
$$
together with the second line of Eq.~(47c), forming the total trace
Hamiltonian and doing a spatial integration by parts, we now get
$$
{\bf H}={\bf H}_{\psi_{(1,2)}}+{\bf H}_{B,{\cal J}_0}+{\bf
H}_{B^\prime,-{\cal J}_0^\prime}~,
\eqno\rm (50a)
$$
with
$$
\eqalign{
&{\bf H}_{\psi_{(1,2)}}=\int~d^3x
{\bf Tr}\left\{-{1\over 2}~\sum_{\ell=1}^3~
\left[\psi_{(2)}^\dagger \gamma^0\gamma^\ell
D_\ell\psi_{(1)}+(D_\ell\psi_{(1)})^\dagger\gamma^0\gamma^\ell\psi_{(2)}\right.\right.\cr
&~~~~~\biggl.\left.-\psi_{(1)}^\dagger\gamma^0\gamma^\ell
D_\ell\psi_{(2)}-(D_\ell\psi_{(2)})^\dagger\gamma^0\gamma^\ell\psi_{(1)}\right]
-m\left[\psi_{(2)}^\dagger
i\gamma^0\psi_{(1)}-\psi_{(1)}^\dagger
i\gamma^0\psi_{(2)}\right]\biggr\}~. \cr}
\eqno\rm (50b)
$$
In Eq.~(50a), ${\bf H}_{B,{\cal J}_0}$ and ${\bf H}_{B^\prime,-{\cal
J}_0^\prime}$ are still given by Eq.~(48b) with the substitution ${\cal
J}_0^\prime\to-{\cal J}_0^\prime$, but now ${\cal J}_0$,
${\cal J}_0^\prime$ are the $0$ components of the fermion source
currents ${\cal J}_\nu,{\cal J}_\nu^\prime$ given in Eq.~(35).

The total trace Hamiltonian dynamics for $\phi$, $\psi_{(1)}$, and
$\psi_{(2)}$ now takes the form of Eq.~(5f), with no further
complications.  For the gauge potentials $B_\mu$ and $B_\mu^\prime$,
however, we encounter the familiar problem that we are dealing with a
constrained system, and so the canonical momenta are not independent.
Focusing henceforth on the potential $B_\mu$ (the treatment of
$B_\mu^\prime$ is completely analogous), we have a primary constraint
$$
p_{B_0}=-{1\over G^2}~F_{00}=0~,
\eqno\rm (51a)
$$
which is satisfied as an identity without use of the equations of
motion.  Differentiating Eq.~(51a) with respect to time, we get the
secondary constraint
$$
0=\dot p_{B_0}=-{\delta\over\delta B_0}~{\bf H}_{B,{\cal J}_0}={\cal
J}_0+\sum_{\ell=1}^3~\hat D_\ell p_{B_\ell}~,
\eqno\rm (51b)
$$
which is the same as the constraint arising from the Lagrangian
equations of motion.  Further time differentiation of Eq.~(51b)
leads to no further secondary constraints, since the equation
$$
\dot{\cal J}_0+{\partial\over\partial t}~\left(\sum_{\ell=1}^3~\hat
D_\ell p_{B_\ell}\right)=0
\eqno\rm (51c)
$$
can be rearranged [with use of Eq.~(51b) and properties of the covariant
derivative $\hat D_\mu$] into the form
$$
\hat D^\mu{\cal J}_\mu=0~,
\eqno\rm (51d)
$$
which is satisfied by virtue of the Lagrangian equations of motion for
$\phi$ or $\psi_{(1,2)}$.  The constraint structure is thus completely
analogous to that of a conventional Yang--Mills gauge field, for which
the simplest way to realize a Hamiltonian dynamics is to use axial
gauge [27,28], in which $B_3$ is taken to vanish,
$$
B_3=0~.
\eqno\rm (52a)
$$
In this gauge we have
$$
\eqalignno{F_{03}
&=\partial_0B_3-\partial_3B_0+[B_0,B_3]=-\partial_3B_0~,\cr
\hat D_3 F_{03} &=\partial_3F_{03}+[B_3,F_{03}]=-\partial_3^2 B_0~, \cr
F_{13} &=-\partial_3B_1~,~~~~~~F_{23}=-\partial_3B_2~,
&(52b)\cr}
$$
and so the constraint
$$
G^2{\cal J}_0=\sum_{\ell=1}^3~\hat
D_\ell\,(-G^2p_{B_\ell})=\sum_{\ell=1}^3~\hat D_\ell F_{0\ell}=\hat D_1
F_{01}+\hat D_2F_{02}-\partial_3^2B_0
\eqno\rm (52c)
$$
can be directly integrated to yield $B_0$ and $F_{03}$, giving (with
$x_3=z$)
$$
\eqalignno{B_0
&=-{1\over 2}~G^2\int_{-\infty}^\infty~dz^\prime |z-z^\prime|({\cal
J}_0+\hat D_1p_{B_1}+\hat D_2p_{B_2})_{z^\prime}~,\cr
F_{03} &={1\over 2}~G^2\int_{-\infty}^\infty~dz^\prime~{z-z\prime\over
|z-z^\prime|}~({\cal J}_0+\hat D_1p_{B_1}+\hat D_2p_{B_2})_{z^\prime}~.
& (52d)\cr}
$$
Substituting Eqs.~(51b) and (52b) back into ${\bf H}_{B,{\cal
J}_0}$, we get
$$
\eqalign{{\bf H}_{B,{\cal J}_0}
&=\int~d^3x{\bf Tr}\,
\left\{-{G^2\over 2}~\sum_{\ell=1,2}~(p_{B_\ell})^2
-{1\over 2G^2}~(F_{03})^2-{1\over
2G^2}~ (\partial_1B_2-\partial_2B_1+[B_1,B_2])^2\right.\cr
&\biggl.~~~~~~~~~~~~~~-{1\over
2G^2}~(\partial_3B_1)^2-{1\over 2G^2}~(\partial_3B_2)^2\biggr\}~,\cr}
\eqno\rm (52e)
$$
with $F_{03}$ given by Eq.~(52d), and so only $B_1,B_2$ remain as
independent dynamical degrees of freedom, with the corresponding
independent canonical momenta $p_{B_1},p_{B_2}$.  It is now completely
straightforward to verify that the operator equations of motion obtained
from the total trace Hamiltonian ${\bf H}_{B,{\cal J}_0}$ of
Eq.~(52e),
$$
\eqalignno{{\delta{\bf H}_{B,{\cal J}_0}\over\delta
p_{B_1}}
&=\partial_0B_1~,~~~~~~{\delta{\bf H}_{B,{\cal J}_0}\over\delta
p_{B_2}}=\partial_0B_2~,\cr
{\delta{\bf H}_{B,{\cal J}_0}\over\delta
B_1} &=-\dot p_{B_1}~,~~~~~~{\delta{\bf H}_{B,{\cal J}_0}\over\delta
B_2}=-\dot p_{B_2}~,~~~~~~&(53)\cr}
$$
are identical to the operator equations of motion obtained from the
total trace Lagrangian.  So we have achieved a consistent Hamiltonian
dynamics.  The generalized Poisson bracket of Eq.~(6a), in axial
gauge, now contains variational derivatives only with respect to the
gluon variables $B_{1,2}$ and $p_{B_{1,2}}$.  We have not verified the
Poincar\'e generator algebra, but just as with the verification of the
Hamiltonian equations of motion which we have described, this should be
a straightforward analog of the conventional Yang--Mills axial gauge
calculation.

At no point in the discussion have canonical commutation relations been
used to get the operator equations of motion.  They have been replaced
in total trace dynamics by the constraints
$$
0={\cal J}_0+\sum_{\ell=1}^3~\hat D_\ell p_{B_\ell}~,~~~~~~
0=\pm{\cal J}_0^\prime +\sum_{\ell=1}^3~\hat D^\prime_\ell p_{B_\ell^\prime}~,
\eqno\rm (54)
$$
and are a quaternionic field theory generalization, for independent
left and right gaugings, of the constraints of Eq. (14b) of Sec. 2.
We conjecture that any operator realization of Eq. (54) gives, via the
total trace
Hamiltonian formalism of Eqs. (47a)--(53), a consistent quaternionic
field dynamics.
\bigskip
\centerline{\bf 5.  Discussion}

In the preceding three sections, we have presented a generalization of standard
quantum mechanics, providing a framework within which one can formulate
quaternionic quantum field theories.  We suggest in the final chapter of [1]
that such theories may play a role in physics between the GUTS scale and
the Planck mass, possibly providing a dynamics for preon models of quarks
and leptons, and that the observed complex quantum field theories associated
with the standard model and its grand unification are effective theories
describing the asymptotic dynamics of the underlying quaternionic fields.

These speculations aside, the concepts of total trace Lagrangian and
Hamiltonian dynamics, and operator gauge invariance, provide rich new
possibilities for the formulation and study of quantum field systems.  Among
the many open questions which remain are:  (1) Can the generalized bracket
of Eq.~(6a) be proved to satisfy a Jacobi identity?  (2) What is the
analog of the Dirac theory of constrained systems for operator gauge
invariant systems?  (3) What is the analog of BRST theory for such systems?
(4) For the gauge models discussed in axial gauge in Sec.~4, what form does the
total trace Hamiltonian take in non--canonical gauges, such as transverse
gauge?  (5) Does total trace dynamics, in the quaternionic case, correspond
to a unitary time development, or is this only a property of complex
quantum mechanics?  (6) Can one usefully characterize the solutions of the
operator constraints of Eq.~(54)?  (7) Given two scalar or fermion
quaternionic operator fields, is there a criterion for determining whether
they are related by a bi--unitary operator gauge transformation?
(8) Can one find a functional integration form of total trace dynamics,
analogous to the Feynman path integral in complex quantum field theory?
(9) More generally, what are useful calculational techniques for the
theories formulated in this paper?
\bigskip
\centerline{\bf Acknowledgement}

I wish to thank many colleagues, cited in full in [1], for helpful
conversations
and correspondence dealing with quaternionic quantum mechanics, quantum field
theory, and related issues.  This work was supported in part by the
Department of Energy under Grant \#DE--FG02--90ER40542.
\vfill\eject
\centerline{\bf References}
\Rf
[1]  S.L. Adler, Quaternionic quantum mechanics and quantum fields
(Oxford University Press, Oxford, 1994).
\Rf
[2]  E. Witten, J. Diff. Geom. 17 (1982) 661.
\Rf
[3]  S.L. Adler, Phys. Lett. 86B (1979) 203.
\Rf
[4]  S.L. Adler, Phys. Rev. D21 (1980) 550.
\Rf
[5]  S.L. Adler, Phys. Rev. D21 (1980) 2903.
\Rf
[6]  D. Finkelstein, J.M. Jauch and D. Speiser, Notes on quaternion
quantum mechanics (1959), published in C. Hooker, ed.,
Logico--algebraic approach to quantum mechanics II (Reidel, Dordrecht,
1979).
\Rf
[7]  G.W. Mackey, Weyl's program and modern physics (1987), in K.
Bleuler and M. Werner, eds., Differential geometric methods in
theoretical physics (Kluwer Academic Publishers, Dordrecht, 1988).
\Rf
[8]  G.W. Mackey, The axiomatics of particle interactions (1992), talk
at the Castiglioncello conference, Int. J. Theoret. Phys. (in press).
\Rf
[9]  A. Connes, Publ. Math. IHES 62 (1983) 44.
\Rf
[10]  A. Connes, Essay on physics and non--commutative geometry, in
D. Quillen, G. Segal and S. Tsau, eds., The interface of mathematics
and particle physics (Clarendon Press, Oxford, 1990).
\Rf
[11]  D. Finkelstein, J.M. Jauch, S. Schiminovich and D. Speiser, J.
Math. Phys. 4 (1963) 788.
\Rf
[12]  L.P. Horwitz and L.C. Biedenharn, Ann. Phys. 157 (1984) 432.
\Rf
[13]  O. Teichm\"uller, J. Reine Angew. Math. 174 (1935) 73.
\Rf
[14]  D. Finkelstein, J.M. Jauch and D. Speiser, J. Math. Phys. 4 (1963)
136.
\Rf
[15]  A. Razon and L.P. Horwitz, Acta Applicandae Mathematicae 24 (1991)
141.
\Rf
[16]  A. Razon and L.P. Horwitz, Acta Applicandae Mathematica 24 (1991)
179.
\Rf
[17]  A. Razon and L.P. Horwitz, J. Math. Phys. 33 (1992) 3098.
\Rf
[18]  P. Cvitanov\'ic, R.J. Gonsalves and D.E. Neville, Phys. Rev. D18
(1978) 3881.
\Rf
[19]  S.C. Lee, Phys. Rev. D20 (1979) 1951.
\Rf
[20]  S.C. Lee, Phys. Rev. D21 (1980) 466.
\Rf
[21]  K.A. Milton, W.F. Palmer and S.S. Pinsky, Phys. Rev. D25 (1982)
1718.
\Rf
[22]  S.L. Adler, Phys. Rev. D17 (1978) 3212.
\Rf
[23]  E. Witten, Phys. Lett. 117B (1982) 324.
\Rf
[24]  S. Weinberg, Gravitation and cosmology (John Wiley, New York,
1972) Sec. 12.2.
\Rf
[25]  S. Ferrara and B. Zumino, Nucl. Phys. B79 (1974) 413.
\Rf
[26]  A. Salam and J. Strathdee, Phys. Lett. 51B (1974) 353.
\Rf
[27]  R.L. Arnowitt and S.I. Fickler, Phys. Rev. 127 (1962) 1821.
\Rf
[28]  A. Hanson, T. Regge and C. Teitelboim, Constrained Hamiltonian
systems (Acad. Naz. dei Lincei, Rome, 1976).
\bye